\documentclass{article}
\usepackage[utf8]{inputenc}

\title{Public-Private Partnership in the Management of Natural Disasters: A Review}
\author{{Selene Perazzini$^{a}$}
\\ \\
\small{$^{a}$IMT School for Advanced Studies Lucca}}
\date{\today}
\usepackage{natbib}
\usepackage[normalem]{ulem}
\useunder{\uline}{\ul}{}

\usepackage{graphicx}
\usepackage{lscape}
\usepackage[flushleft]{threeparttable}
\usepackage{adjustbox}
\usepackage[labelfont=bf]{caption}
\usepackage{booktabs}
\usepackage{siunitx}

\usepackage{dcolumn}
\usepackage{longtable}
\usepackage{threeparttablex}

\usepackage{longtable,booktabs,threeparttablex}

\begin{document}

\maketitle

\begin{abstract}
Natural hazards can considerably impact the overall society of a country. As some degree of public sector involvement is always necessary to deal with the consequences of natural disasters, central governments have increasingly invested in proactive risk management planning. In order to empower and involve the whole society, some countries have established public-private partnerships, mainly with the insurance industry, with satisfactorily outcomes. Although they have proven necessary and most often effective, the public-private initiatives have often incurred high debts or have failed to achieved the desired risk reduction objectives. We review the role of these partnerships in the management of natural risks, with particular attention to the insurance sector. Among other country-specific issues, poor risk knowledge and weak governance have widely challenged the initiatives during the recent years, while the future is threatened by the uncertainty of climate change and unsustainable development. In order to strengthen the country's resilience, a greater involvement of all segments of the community, especially the weakest layers, is needed and the management of natural risks should be included in a sustainable development plan.
\end{abstract}

\section{Introduction}
Natural risks pose a broad range of social, financial and economic issues, with potentially long-lasting effects. Historically, governments have mostly addressed the financial effects of natural events on an ad-hoc basis, but, since the 1970s, 
a strong awareness-raising activity by the United Nations has profoundly changed the attitude of the national authorities.\footnote{
The United Nations dates the beginning of the global disaster risk reduction process to the International Expert Group Meeting in July 1979, but the first International Framework of Action - the international Decade for Disaster Reduction - began 10 years later, on January 1990. Then framework has then been followed by the Yokohama Strategy in 1994. In 1999 the United Nations established the UNDRR (United Nation office for Disaster Risk Reduction), a secretariat dedicated to facilitate the implementation of the ``International Strategy for Disaster Reduction''. The office presented a first plan - the ``Hyogo Framework for Action 2005-2015'' \citep{Hyogo} - to explain, describe and detail the work that is required from all different sectors and actors to reduce disaster losses. When the framework reached maturity, a 15-year-long, voluntary, non-binding agreement aiming at ``\textit{the substantial reduction of disaster risk and losses in lives, livelihoods and health and in the economic, physical, social, cultural and environmental assets of persons, businesses, communities and countries}'' - the ``Sendai Framework'' \citep{Sendai} - was adopted by UN Member States. A series of agreements - the 2030 Agenda, the Paris Agreement on climate change, the New Urban Agenda, the Addis Ababa Action Agenda and the Agenda for Humanity - then sought to coordinate risk reduction with the other global challenges. Other international organizations are also involved in similar projects for the prevention, mitigation and management of natural disaster risk, such as ADB (Asian Development Bank), FAO (Food and Agriculture Organization of the United Nations), OECD (Organization for Economic Co-operation and Development) and the World Bank. For example, in 2012, G20 Finance Ministers and Central Bank Governors, along with G20 Leaders mandated the OECD to develop a voluntary framework for strengthen disaster risk assessment and financing \citep{OECD1}.} According to the United Nations, all governments should not just guide citizens toward recovery in the aftermath of an event, but prepare them and create the conditions necessary to ensure rapid resilience. To this aim, OECD, G20 \citep{OECD1}, the World Bank and GFDRR \citep{WB}, encourage countries to adopt a comprehensive disaster risk management strategy, which should be articulated in a series of coherent and coordinated actions, well distributed and defined over time and aimed at addressing a specific phase of the disaster. In particular, the strategy must include risk assessment, risk reduction, preparedness, emergency response, and recovery.

Among these phases, all equally important, risk reduction is by far the most complex to plan. Many tools able to reducing the financial impact of natural events exist, and can be divided into three categories based on how the risk is addressed. We distinguish between risk mitigation, risk retention and risk transfer. Mitigating risks means acting on the physical and environmental conditions responsible for the financial impact, therefore all structural interventions aimed at decreasing the probability of an event occurring (e.g. reservoirs), the vulnerability of exposed assets (e.g. retrofitting on private homes) or the number of goods and individuals exposed (e.g. restricting building permission in high-risk areas) fall into the first category. Since risks cannot be completely mitigated and structural interventions might not be cost-effective \citep{Hudson}, a good risk management strategy should always include some degree of financial protection \citep{WB}. Financial instruments\footnote{Two approaches to risk financing exist and correspond to different financial instruments. Risks might be addressed ex post by means of existing resources and powers, or ex ante with financial mechanisms explicitly arranged or secured beforehand. For example, ex ante instruments that governments can rely on are reserve funds, contingent credit facilities, re/insurance, catastrophe-linked securities. Examples of ex post financing are budget reallocation, debt financing, borrowing and taxation. Note that ex post financing does not preclude the establishment of institutional arrangements that specify, ex ante, the government’s financial commitments \citep{OECD1}.} for risk reduction can distribute the costs over time by accumulating sufficient capital to face the expected losses of future events (risk retention), or transfer the risk to specialized subjects, i.e. insurers and reinsurers, or to the market through catastrophe linked securities. Both risk retention and transfer facilitate emergency response and speed up recovery by providing resources in the immediate aftermath of an event.

The governments not only should select and adopt the most suitable risk reduction measures, but must also ensure that individuals have access to them. Since private insurance on buildings and/or on other movable assets is the main risk financing tool for businesses and households, the \cite{OECD1} recommends that governments \textit{``assess their availability, adequacy and efficiency to the population and within the economy, as well as their costs and benefits relative to other types of possible risk reduction measures''}. Nevertheless, a series of market failures endanger financial and insurance markets\footnote{Main market failures in disaster risk management relate to the insurability of risks, information asymmetry, adverse selection, consumer behavior, moral hazard and charity hazard. As far as insurability concerns, spatial correlation among insured assets constitutes a central issue for disaster management because generates the potential for enormous losses to the insurers \citep{Glauber}. For example, a series of hurricanes in the US during the 1990s led to a consistent number of insolvencies \citep{Matthews,Mills}. As a consequence, insurance included higher risk-load in premium rating for high-risk areas \citep{Feldblum,Kreps,Meyers,Mango97,Mango98,Kreps2}, that often do not meet the demand from rational purchaser \citep{Kousky}. Along with behavioural bias \citep{Kunreuther}, climate change further complicates the development of financial and insurance market. The \cite{GenevaAssociation} warns that return periods and correlation among claims for several high-loss extreme events are ``\textit{ambiguous rather than simply uncertain}'', and raises concerns about the future sustainability of insurance business on natural risks. Failures in capital markets have been explored by \cite{Froot}, that found that securitization is not always the lowest-cost way to transfer risk due to supply restrictions associated with capital market imperfections and market power exerted by traditional reinsurers.} and makes it necessary to further reduce the risks before transferring them. In this respect, governments have often intervened by investing in risk mitigation and increasing public awareness among the population.
In some cases, when the costs of policies were prohibitive for some segments of the population, the authorities directly intervened in the insurance market by establishing a public-private partnership. Though in some circumstances efforts of governments have already been substantial, the \cite{GenevaAssociation2} expects the role of the insurance industry to become increasingly relevant in the next future and urges governments to increase their commitment in monitoring socio-economic risks of climate change, developing risk management plans for all sectors of the economy, and establishing relevant public-private partnerships with insurers to enhance socio-economic resilience.

The fragility of the insurance industry is only part of the problem facing countries. Governments themselves are in fact significantly exposed to disaster risk: public exposures are large, including human losses, injuries, damage to public goods, tax pressures resulting from financial commitments and unplanned post-disaster financial assistance, as well as potentially negative changes in macroeconomic conditions such as possible lower economic growth or lower tax revenues.\footnote{Losses that the government may incur can be both explicit or implicit: the expenses that could derive from the reconstruction of public goods and infrastructures or other financial commitments following a disaster are explicit; on the contrary, expenses that do not reflect any type of commitment or liability, but which can still occur due to a perceived obligation are implicit.} In order to protect the national financial stability, the Sendai Framework claims that the government, while guaranteeing social assistance, should share responsibilities with private stakeholders and therefore private initiatives in prevention and financial protection should be encouraged. As emphasized by the \cite{OECD2}, improving public awareness reduces the human-induced factors that make a major contribution to the cost of disasters and alleviates losses on public finances. However, educating and informing the society is usually not enough. 


Hence, if some degree of government involvement is necessary to protect the most vulnerable layers of the society and the market, how can authorities balance public and private initiatives? According to \cite{Jaffee}, public initiatives should only complement private activities and the role that the government should assume depends on the relationship between objective and subjective probabilities of loss. In case of perfectly rational individuals with objective perception of risk, an active role of government is necessary during emergency response only; if individuals underestimate their risk, investments in public awareness are needed or, alternatively, mandatory insurance purchase might be introduced. Whether other differences between objective and subjective probabilities are not generated from behavioural biases, the government should identify and implement the solution that addresses the specific market failure in the most efficient way.

Unfortunately, identifying market failures is complicated. To make matters worse, ``\textit{with increasing complexity and interaction of human, economic and political systems within ecological systems, risk becomes increasingly systemic}'' \citep{GAR2019}, and responsibilities increasingly blurred. In increasingly uncertain and complex contexts, cooperation between all the subjects involved - individuals, businesses, authorities - is essential to build the community's resilience. Although most countries are still not adequately prepared to deal with the consequences of possible future disasters\footnote{According to the report by \cite{Wilkinson2}, a number of disaster risk management activities were conducted during the past decades (most of which were relatively low cost), but, however, they were not as effective as they could and disaster losses increased during the Hyogo Framework of Action.}, some have already implemented sophisticated risk management plans which envisage a public-private partnership with the insurance sector. Almost all of these few virtuous countries have intervened in the insurance sector, becoming insurers, reinsurers or, in the poorest economies, by activating micro-insurances. These partnerships are then supported and strengthened by governments through a series of legislative provisions and investments, which make each strategy unique. Although it is not possible to replicate any of these strategies in other countries, some useful lesson can still be drawn from them. In the next section the main public-private partnerships currently in force are analyzed. The benefits of these partnerships are widely recognized, but some important weaknesses have also emerged, which we discuss in section \ref{Failures}. Among these, risk understanding and government's attitude toward natural risks are today the two major limits in disaster risk management. Many of these weaknesses can be overcome by adopting an even more inclusive approach, which involves a greater number of subjects and therefore allows to monitor the risk on the whole society. For these reasons, \cite{GAR2019} argues that countries should move towards a community-based approach to risk management, and section \ref{CBA} deals with this. To conclude, section \ref{Future} discuss the next challenges in disaster risk management.


\section{Public-Private Partnership in Insurance}
There is a widespread agreement on the benefits of public-private partnerships for the management of natural disasters \citep{Kunreuther2006,WB11,IPCC2} and in particular, public intervention in the insurance sector is increasingly proving to be effective, especially in the poorest countries. Government-supported initiatives are in fact able to distribute risks and losses over the entire population and over time \citep{KunreutherPauly3}, and are much more flexible than private solutions as they are not tied to profit objectives \citep{Penning-Rowsell}. When insurance schemes are properly designed and supported, they communicate risk to the population, foster adaptive responses and risk reduction and above all improve economic stability and protect the well-being of the community \citep{KunreutherPauly3,Lotze-Campen,Hudson,KunreutherLyster,Kousky2018,Linnerooth-Bayer}. As argued by \cite{Bruggeman}, however, public intervention are beneficial only if they solve a specific market failure that the private sector is not able to cope with on its own. Otherwise, the State's entry into the insurance (or reinsurance) market might play a distorting effect. Unfortunately, identifying and recognizing the market failure may not be easy. 

Provided that the public intervention is necessary, the effectiveness of the insurance system depends on a number of conditions. First of all, it is essential to achieve a satisfactory understanding of the natural phenomenon and the extent of the losses to which it can lead. In particular, capturing the spatial correlation that binds insured properties is fundamental as it challenges the rating process and, in turn, the financial stability of the system, as happened for example in the U.S. corn insurance market \citep{Woodard}. Secondly, the business should be supported by a coordinated set of actions aimed at overcoming all the frictions that generate low take up rates, such as lack of trust in the institution, liquidity constraints, and limited salience among citizens \citep{Cole3}. Educating the population has often fostered the adoption of policies \citep{Bogale,Gan}, and income and development support measures proved effective in some circumstances \citep{Greatrex, McIntosh}. If this is not sufficient, mandatory insurance purchase tackles the root problem \citep{KunreutherPauly3}, though this solution may not be well received by citizens. Furthermore, raising awareness of the population on natural disasters in quiet periods is also decisive, as the prolonged absence of major events leads to lowered attention and decreases policy's purchase \citep{Gan, Gallagher}. 

In this section we present and discuss the three fundamental types of public-private-partnership in the insurance business - public insurance, public reinsurance, and micro-insurance -, how governments have implemented them and the difficulties they encountered.

\subsection{State-owned Insurance}

When a peril has a potential high economic impact in a given area, insurance companies may fear for their financial stability and therefore decide to limit the offer or to provide coverage at an excessively high price in that area. If the area is large and the number of individuals and uncovered properties is high, the lack of insurers becomes a huge problem for the government as citizens might demand for public intervention in the aftermath of an event. When facing this situation, some countries reputed offering policies at a price affordable for all citizens more efficient than deploying capital in ex post relief programs \citep{KunreutherPauly3} and have therefore established a state-owned insurance company. As public companies aim at solving gaps, coverage is provided for perils that cannot be borne by private companies only. In fact, most of the schemes deal with a single peril, such as earthquakes in California or Turkey, or with a restricted set of them. Insurable items are also limited: coverage is usually provided only for buildings, sometimes even for vehicles, while other risks, such as business interruptions, are rarely covered.

Table \ref{tab:State-owned Insurance} describes the main public insurance against natural disasters currently in place. Most of these companies were born out of heated debate between insurers operating in the area and the national authorities, and are supported by both. Sometimes the role of the private sector is limited to technical advice during the creation of the public program and to data provision, other times the public company shares the risks with the private ones through co-insurance. Furthermore, private companies almost always act as intermediaries between citizens and the public company by underwriting policies and transferring risks and the related premiums to the state-owned entity in exchange for a fee. Governments typically support the public company by assuming the role of guarantor in exchange of a charge (e.g. EQC in New Zealand) or for free (e.g. Consorcio de Compensacion de Seguros in Spain). Alternatively, governments may provide a prearranged facilitated access to credit, as in the case of the National Flood Insurance Program (NFIP). In addition, State-owned companies may also benefit of contingent credit lines from international organizations such as the World Bank.

Since the main purpose of government pools is to give the chance to most of the population to get insured, the policies are sold at a low price or at least at a premium lower than that offered by private companies.
This choice has two important drawbacks. First, the low premiums of the public insurer can compete with the few private companies that have decided to offer the policy, generating a crowding out effect and weakening the private sector \citep{McAneney}. Second, if low rates are not actuarially sound, they expose public insurers to a risk of reserve depleting higher than that legally allowed to traditional insurers. In addition, governments often apply flat premium rates that include subsidy among individuals, but fail to create risk-reflecting reserves. Low rates therefore make government guarantees or other forms of public financial support necessary to the sustenance of the program but, in order to limit public capital injections, the pool should ideally become self-sustained at a certain time. To this aim, governments can reduce the risk beared by the company by encouraging and committing communities to risk mitigation \citep{Kunreuther2,Kunreuther3}. Building codes and premium discounts for properties subjected to structural strengthening interventions have been extensively adopted, albeit with varying results. The risk reduction is in fact largely demanded to citizens, who may consider the investment not advantageous \citep{Kleindorfer2}, and therefore countries that have applied more binding risk mitigation plans have obtained greater participation.

The company's financial exposure can then be reduced by applying deductibles to the policies or transferring the risks through reinsurance. Both components are important: the former preserves a certain degree of citizens' responsibility; the latter allows the State-owned company to get rid of the higher losses by transferring them to specialized bodies or on international markets, so as not to unduly affect the insurance reserves and public resources following a catastrophic event \citep{OECD4}. In particular, as risks evolve rapidly and disasters become more and more frequent, reinsurance is essential for public companies to survive \citep{Seo}.

Although all the schemes in Table \ref{tab:State-owned Insurance} are profoundly different from each other, any of them requires four elements to properly function: strong government financial support, a great commitment to risk reduction, high citizen participation and ongoing access to the reinsurance market. Since State-owned insurances are just a few and some of them are extremely young, it is difficult to outline how to properly balance the four components. However, the history of NFIP, which is one of the oldest and most studied public insurance programs, has shown how decisive these components are.\footnote{NFIP is a public insurance program for flooding in the U.S. and provides policies only if the community joined a floodplains laws and ordinances program. The program applies risk-based premiums (although some subsidy was introduced in 2014 with the Homeowners Flood Insurance Affordability Act) defined on the flood maps. In order to better capture the real risk of the different areas, the government committed to update and complete the flood risk maps. The program has been hardly criticized because rates did not provide the necessary income to build long-term reserves \citep{USAccounting} and flood mapping process was not so effective and timely as planned \citep{KouskyKunreuther}. In addition, NFIP was not authorized to secure private reinsurance until the Biggert-Waters Flood Insurance Reform Act of 2012 and the Homeowners Flood Insurance Affordability Act of 2014. Although rates have been updated several times, the program reached the statutory borrowing cap from the Treasury in September 2017. The U.S. Congress cancelled more than half of the debt, but NFIP had to borrow additional capital two months later. The company's outstanding debt is currently \$20.5 billion and is likely to grow as its annual probable maximum loss is over \$40 billion and its capacity to pay claims without borrowing is \$5.4 billion \citep{NFIP}. As extreme rainfall along the Gulf Coast are on the rise \citep{van_Oldenborgh_2017}, it therefore seems very unlikely that the program will be able to resist without the strong support of the State. Moreover, unless reauthorized or amended by the Congress, the NFIP's borrowing limit from the Treasury will be reduced from \$30.425 billion to \$1 billion at the end of September 2020. Although late access to reinsurance and low premiums have certainly played a role in the NFIP's debt accumulation, part of the financial weakness of the program is due to the insufficient participation of the communities. In general, individuals have not invested enough in risk mitigation and the government has not been able to identify the right stimuli. A more frequent remapping and the consequent adjustment of the premiums could have helped to spread awareness and encourage private investments \citep{Michel-Kerjan4,KouskyKunreuther}. Along with insufficient risk mitigation, NFIP has also recorded low take up rates \citep{Dixon}. }

Voluntary citizen participation is the most common issue in natural disaster insurance and, despite favorable cost, most State-owned insurances record low take up rates also. Undesired consequences of low insurance penetration are potentially high costs of government's post-disaster assistance \citep{Dixon}, and reduced access to the reinsurance market for the insurer \citep{vonLucius}. There are several reasons why the population exhibits careless behavior towards natural disasters' prevention, but state-owned insurances seem particularly affected by charity hazard \citep{WB9,Raschky,Basbug,Marshall}. The belief that the government will help irrespective of owning an insurance policy is stronger when governmental relief is more certain \citep{Raschky2}, and States offering a public-private insurance have usually been very generous to the community when hit by a disaster. Along with charity hazard, low risk perception and poor policy understanding are also quite frequent \citep{Chivers}. Introducing mandatory insurance purchase can rise take up rates \citep{KunreutherPauly3,Kriesel}, provided that the obligation is properly formulated and monitored \citep{Dixon}. For example, in Turkey, property-owners are required to prove to have valid policy only when they want to buy or sell a house or to obtain a new account for water and electricity services. As argued by \cite{Basbug}, this sporadic check does not enforce ongoing renewal of the insurance. The Turkish government has therefore activated some initiatives aimed at promoting awareness, many of which have been designed so that the most sensitive citizens involve an increasing number of acquaintances. For example, a 20 percent premium discount is offered if eight individual apartment unit owners from the same apartment complex jointly take out a policy, and is supporting the growth of a large volunteer and civil society network \citep{Basbug,WB10}. As frequent monitoring mandatory requirement might be expensive, extending the policy validity to multi-years might also help \citep{Michel-Kerjan3,Kleindorfer}.

Along with scarce participation of the citizens, State-owned companies should also deal with all the problems traditionally affecting insurers. First of all, adequate risk assessment might be extremely challenging where historic losses are not available. An insurance scheme might be hence designed based on simulation techniques, as in the case of the Turkish Catastrophe Insurance Pool \citep{Linnerooth-Bayer2}, but this brings additional uncertainty to the estimates \citep{Basbug, Cakti}. The sensitivity of the cost of reimbursements to changes in market prices or the reconstruction of homes should also be taken into account when constructing reserves. For example, in New Zealand, the State-owned insurance EQC operates in co-insurance with the private sector and pays its reimbursement quota to the insured along with the traditional insurer. After the Christchurch earthquake sequence in 2011, different private companies met their obligation at different times (some even took more than 5 years) and, since the reimbursement provided by EQC is equal to the cost of rebuilding the home, the cost of the event on the company's reserves was strongly affected by the market price variation of both construction works and materials \citep{Wood}. At last, State-owned insurance companies face additional transaction and administrative expenses compared to private companies, which can also be very high. \citep{Marshall,Michel-Kerjan4}. 



\subsection{Public Reinsurance}

Some countries faced great natural risks with an underdeveloped insurance industry that was not solid enough to manage the potential losses of the entire population. In order to strengthen the private insurance sector and foster its growth, a few governments established a State-administered reinsurance company to which insurers can or must transfer natural risks undermining their financial stability (some examples are reported in Table \ref{tab:State-owned Reinsurance}). Compared to providing a State guarantee to insurers, this strategy preserves individual responsibilities and requires the participation of all individuals who contribute to the creation of the financial exposure as the cost of reinsurance is charged on the final price of the policy \citep{Bruggeman}. 

The characteristics and problems of public reinsurance are very similar to those of State-owned insurances. First, the premium paid for reinsurance is typically low because the authorities want the insurers to keep citizens' premiums low. However, this solution is disproportionately beneficial for insurers which, unless otherwise required by law, are free to apply the rating mechanism they want. 
For this reason, some government-reinsurers fixed the property-owners' premiums by legislative decree and/or obliged insurers to offer the policy. In particular, policies for natural disasters are often compulsorily included in other basic policies, the purchase of which may be mandatory for property-owners. In general, this has led to satisfactorily high insurance penetration rates. However, low reinsurance premiums may also challenge the Company's financial stability, as happened to the French Casse Centrale de Reassurance (CCR) in 1999.\footnote{In France, the publicly owned reinsurance company Casse Centrale de Reassurance (CCR) hit bankruptcy in 1999 due to too low fees and too much confidence in the unlimited State guarantee \citep{Vallet,Bruggeman}. Measures were then taken which changed the conditions of the subscription. There is also concern for the young Flood Re in the UK, which seems to generate strong pressure on public finances \citep{Surminski}.}

In order to ease financial pressure, governments often include risk mitigation in the State's risk management plan, although the measures envisaged have not always been effective. In addition to the difficulties in incentivizing private investments in risk reduction, in many countries the greatest obstacle to risk reduction activities have been the government's own management objectives. In France, for example, when the mandatory insurance requirement came into force and the CCR was instituted, the flood risk mitigation measures required clashed with the growth interests of the local authorities and consequently were not implemented properly \citep{Vallet}. In the UK, instead, the government does not seem to actually encourage risk reduction despite it is required by the agreement signed with the insurers \citep{Surminski2,Penning-Rowsell,Surminski}. In Florida, the Florida Catastrophe Insurance Fund was launched to encourage urban growth but, according to \cite{Seo}, this increased the risk exposure over time powered by climate change. In 2009, to decrease exposure, the government has hence activated a program for gradual privatisation of its risk, which also introduced the adoption of retrocession lines and insurance-linked securities (ILS). In this respect, retrocession is proving increasingly important to ensure the continuity of public reinsurance
due to climate change and the slow progress in risk mitigation.

To date, the strength of probably the most stable public reinsurance system in place, the Japan Earthquake Reinsurance Co., is due to a combination of all the revised elements: an adequate risk preparedness and mitigation, a strong political leadership, a structured risk retrocession plan, and the simplicity of the policies that has favored adoption by citizens \citep{Takeda}.




\subsection{Micro-insurance}

In the poorest countries, natural disasters generate far more complex social issues than in developed ones, such as malnutrition, school dropouts, increased poverty. Moreover, the risk of natural disasters can slow down the development of these countries, as farmers are more reluctant to invest in new cultivation techniques that can potentially boost productivity in the long run but might generate greater loss in case of a disaster \citep{WB}. In these contexts, insurance can help safeguarding productivity and, potentially, economic growth. However, the poorest have hardly access to risk transfer.

The reasons for the low (or absent) penetration of insurance markets in this segment of the population are manifold and are only partially related to the supply side. In fact, if on the one hand the insurers often fail to assess the risk in these territories due to the lack of historical data, on the other the individuals are very reluctant to purchase. Several studies on African and Asian regions have in fact brought to light a strong distrust of population towards the companies that offer insurance or the institutions that support them \citep{Cole4,Greatrex}.

In this complex context, international organizations, especially the World Bank, and local or national governments have begun to adopt several form of micro-insurance, especially index-based insurance. An index-based (or index-linked) insurance is an insurance policy whose payout is triggered by an easily-measured event, represented through an index, typically concerning weather conditions (e.g. rainfall below a certain level). Index-based insurance are usually applied to agricultural risks and many initiatives have been activated by local governments in several regions of Africa and Asia. A great advantage of these tools is that they do not need information on individual losses, but only weather or environmental data which are more easily available and less expensive to monitor. For the same reason, the compensation mechanism is far more transparent, thus preventing moral hazard and facilitating the access to reinsurance market for the insurer \citep{Alderman, Cole4}.

The main issue related to index-linked insurance is basis risk: it is possible that an individual receives a payment when he has not suffered losses or that he does not receive it against a large loss. In this regard, various solutions for fine tuning indexes have been proposed in the literature, the main ones being the use of early warnings and seasonal forecasting in the payout triggering mechanism and the definition of complex indices incorporating multiple climatic measurements and built on a better geographical granularity \citep{Rao,Daron,Dercon,Conradt}. 

On the demand side, however, frictions are related to non-economic factors, including levels of financial literacy, liquidity, distrust \citep{Cole4,Eling,Bogale,Greatrex}. \cite{Cole4} found a positive effect of involving non-governmental organisations on take up rates, though the impact differs depending on their reputation. The organisation staff might help overcoming distrust and spreading knowledge of the products. Investing in financial literacy and training courses is also highly recommended \citep{Cole4,WB6}, though empirical evidence is confusing \citep{Binswanger-Mkhize,Cole4,Bogale}.

In addition to the difficulties in defining the instruments and the poor grip on the demand, other factors challenge the future of index-linked securities. First, many of these initiatives are supported by the authorities through vouchers or remittances and this support is essential to allow the establishment of the insurance scheme. However, if the system fails to develop properly and does not become profitable to the provider, it is likely that the insurer will stop offering the policy \citep{Alderman}. Further uncertainties concern the future affordability of the policies. \cite{Siebert} notes in fact that the premiums in the Sahel region are extremely sensitive to the climatic parameters of the model and this could lead to a considerable increase in prices in high-risk areas - an increasingly likely scenario, given the strong effects of climate change on weather events. Hence, in order for micro-insurance to continue, an effort from the public sector is still needed \citep{Cole2}. Governments will have both to invest in facilitating access to risk transfer to the poorest, and to incentivize companies to keep offering policies and innovate index-linked insurances.


\section{Failures}\label{Failures}
Some theoretical study show that public-private partnerships are able to offer a more efficient level of natural-catastrophe insurance than free markets \citep{Burby,Charpentier} but a series of deficit characterizes the history of government pools. In this respect, \cite{McAneney} argues that ``\textit{Government pools usually contain an inherent contradiction in trying to provide low cost insurance to high-risk properties and so the funding of deficits to which they are inevitably prone becomes important}''. Fat-tails and spatial correlation make aggregate losses extremely volatile and also contribute to insolvencies \citep{Kousky2,Kousky}. Moreover, most of the government supported insurance apply flat premium rates that do not reflect the asset's riskiness. This choice may be motivated by economic arguments \citep{Hallegatte}, but might fail to provide the necessary income to build long-term reserves.

Public-private insurers can in principle minimise losses by encouraging risk mitigation \citep{Kunreuther2,Kunreuther3}, but little evidence about this can be found in practice. In particular, insurance companies try to encourage risk mitigation by offering discounted policies. However, in order for this initiative to be successful, insurers must have the opportunity to apply risk-based premiums \citep{Kunreuther3} or, if the government demands a form of subsidy in the rating, the authorities must actively engage in risk mitigation, for example by outlining well-enforced building codes that force property owners to adopt cost-effective protective mechanisms \citep{KunreutherWB}. However, defining effective rules on construction and policy purchase and enforcing them has often proved practically difficult and not always in line with the other management objectives of the local authorities.

Alternatively, strengthening the business by adopting new forms of reinsurance coverage can protect insurers against potential insolvency from disasters too \citep{Kunreuther,LeeYu} and for this reason the ILS market has significantly grown in the last decades \citep{Cummins1,Cummins3,CumminsBarrieu}. Despite this, the ILS market is still relatively young and countries are often reluctant to adopt these securities because they usually lack experts who can oversee their construction and issuance \citep{OECD3, Michel-Kerjan}.\footnote{In order to incentivize the adoption of catastrophe-bonds, in 2009 the World Bank has launched the MultiCat program, in which it offers its technical support and act as an arranger. In the same year, Mexico benefited from the program and issued a US\$290 million cat bond with a three-year maturity \citep{WB14}.} In addition to lack of technical knowledge, the development of the ILS market could also be hindered by the crowding out effect\footnote{The crowding out effect of public programs refers to all those situations in which government-supported initiatives meant to cover the uninsured prompt those already enrolled in private insurance to switch to the public program.} that governments involved in the free market exert on more efficient private reinsurance solutions \citep{Cummins2}.

According to \cite{GAR2019}, to date countries have ``patchy'' implemented their risk management strategies. Most are addressing the consequences of disasters rather than trying to reduce their actual risk. The great weaknesses of the national strategies currently in place and the reasons that led to a too slow development of proactive strategies at the national level can all be traced back to two major problems: insufficient risk understanding and weak governance \citep{Opitz-Stapleton}.

\subsection{Risk Understanding}
The effectiveness of risk management policies strongly depends on the ability to identify and assess the risks \citep{UNISDR}. Knowing the probability of occurrence of the events and their potential impacts\footnote{When assessing natural risk a variety of impacts should be considered on different groups across the society. Impacts might be direct or indirect, as shown by \cite{WB} for the government, homeowners, farmers and the poorest.} constitutes the basis for developing and evaluating the whole range of risk management strategies, such as emergency plans or cost-benefit analysis\footnote{As argued by \cite{IPCC1}, use and applicability of cost-benefit analysis to risk reduction measures are constrained by important limitations, that \cite{Mechler} summarizes in: (i) representing disaster risk, (ii) assessing intangibles and indirect benefits, (iii) assessing portfolios of systemic interventions versus single interventions, (iv) the role of spatial and temporal scales. Despite these limitations, the author argues that cost–benefit analysis remains an important tool for prioritizing efficient disaster risk measures and is well suited for the evaluation of infrastructure-based options. By contrast, preparedness and systemic interventions can be better evaluated by means of other tools such as cost-effectiveness analysis, multi-criteria analysis and robust decision-making approaches.} of risk reduction measures. 
It also allows the decision-making process to develop skills to be adapted to local risk profiles and the social conditions of the communities involved, promoting awareness of potential risks among the society. A well-established collection of data on risks, exposures, vulnerabilities and expected losses is fundamental for the success of any risk management strategy \citep{KunreutherWB}.

Both the Hyogo and the Sendai Frameworks have underlined the importance of risk understanding and, in turn, of data collection and risk assessment. Since then, substantial progress was made, but, nevertheless, major gaps still affect many countries \citep{UNISDR}. In particular, \cite{GAR2019} identifies four challenges about data on risks: availability, quality, accessibility and application. Availability concerns data collection, a necessary step for risk assessment. Understanding natural risks requires an enormous amount of information that are costly to collect and, in addition, natural disasters are rare events and therefore creating a database requires time, at least decades. Along with long times of observation, high-quality data is necessary to guarantee effective analysis. Insurers and reinsurers are among the major data producers in the world in terms of both dimensions and quality of their databases, but they are usually not keen to share their database. Data accessibility strongly limits the analysis of natural riskiness, and is not confined to private entities as several countries show difficulties in data-sharing among government institutions. However, accessing to a high-quality database is not sufficient for accurate analysis: data should in fact fit the purpose of the study. 

Data analysis is by far more problematic than data collection. As risk is determined by a combination of hazard, exposure and vulnerability, any change in the society, landscapes, or technology might completely reshape the area's risk profile. As a consequence, past events might no more be representative of the area and this questions the possibility of projecting the future from the past, or at the least, that classical statistical techniques can be used.
To overcome these problems, a new approach to risk assessment,  called ``catastrophe-modeling'', has been developed.\footnote{Catastrophe-modelling began in the 1960s, but its adoption is much more recent. The first commercially-produced model dates back to twenty years later. When introduced, their use was not widespread. They became increasingly popular in the insurance industry from 1989, when Hurricane Hugo and the Loma Prieta Earthquake caused severe losses to US insurers, and then Hurricane Andrew in 1992, that led nine insurers to insolvency, furthermore incentivized their adoption \citep{Grossi}. Since the early beginning, insurance industry has been the most important driver of their development. From the 2000s developers have started including insurance pricing in the models, and actuarial standards and guidelines for the use of catastrophe modelling have been published by the actuarial society of both Europe and the US \citep{Mitchell-Wallace}.} Catastrophe models are softwares that combine geological, engineering, IT and statistical knowledge to simulate the effects of natural events on the territory.\footnote{Note that, despite these models simulate the impacts of natural disasters to overcome lack of data, historical records are still necessary and serve as input for the simulation process.}
These sophisticated tools are widely adopted by the insurance industry and are increasingly used by governments also.\footnote{The first country in investing in catastrophe-modelling was the US. In 1997 FEMA produced HAZUS, a catastrophe-model to estimate earthquake losses in the Country. The model has later been extended to floods and hurricanes. More recently, around the 2010's, there has been a large-scale revisions of the existing catastrophe-models and new ones emerged. Some of these have been produced by governmental organizations, like for example, R-FONDEN in 2007, created through a partnership between the Mexican Natural Disasters Fund (FONDEN) and the Ministry of Finance with the technical support of the Institute of Engineering of the UNAM (Universidad Nacional Autónoma de México) \citep{FONDEN}.} These softwares require continuous updating and their developers are constantly striving to achieve ever more accurate predictions of losses. To date, there are models capable of describing almost all natural risks, but some important eventualities are not yet satisfactorily represented, including cascading effects, multi-hazard analysis, spatial correlation between assets exposed \citep{Mitchell-Wallace}. Furthermore, these models are tied to the characteristics of the areas on which they are built and might not represent regions that are too different from the reference area.

\subsection{Governance}

Currently most countries are addressing the consequences of disasters rather than trying to reduce their actual risk \citep{Opitz-Stapleton, GAR2019}. This purely ``corrective'' attitude\footnote{According to \cite{Twigg}, corrective risk management are project-oriented strategies composed by measures that address specific current risks only. As opposed to the corrective strategy, the progressive disaster risk management is process-oriented and builds a range of capacities to cope with future threats, both anticipated and unforeseen.} is the main weakness of disaster risk management strategies, which should instead define a plan of interventions aimed at building the skills necessary for the community to face adverse events or to limit them. However, developing ``progressive'' disaster risk management is much more complex than just dealing with consequences. First of all, managers should move from a single-threat to a multi-threat perspective, recognize the existence of multiple sources of risk, potentially correlated, and prepare to face all of them \citep{GAR2019}. Secondly, the manager must prepare a plan of interventions that should be effective in the long run. For example, in the case of risk mitigation interventions, this means not only to fund the initial investment, but also to provide for all maintenance activities that will guarantee its future functioning. Finally, it is essential that the risk management strategy is included into a broader management plan, so that consistency between the actions of the various governmental offices will be guaranteed. For example, it urban development plans should not provide for settlements expansion in high-risk areas \citep{Rozenberg}.


This approach is much broader and also much more expensive in terms of both time and cost than simply fixing the consequences. In addition to the difficulty of planning, other factors negatively influence managers' choices. Studies on the relationship between natural disasters and election suggest that it may not be convenient for a government to invest in disaster reduction. In fact, if on the one hand the population tends to blame the government for natural disasters \citep{Achen}, on the other hand politicians are discouraged to invest in risk mitigation because electorate has short memory and benefits from the intervention may take years and may even appear once the ruling party has changed \citep{Cavallo}. In addition, \cite{healy_malhotra_2009} found evidence that voters reward presidential party for disaster relief but not for investing in disaster preparedness. This might explain why governments appear so generous in spending on disaster relief during the election \citep{Cole}. Along with political pressure, decision-makers might fail to recognize the importance of hazards and vulnerability to national development, might be excessively risk-prone or reluctant to allocate substantial resources for events that might not even happen \citep{Michel-Kerjan2,Opitz-Stapleton}.


\section{Toward a Community-Based Risk Management Approach}
\label{CBA}

As countries slowly revise their plannings, risk evolves quickly: the sheer number of people on Earth, climate change and the dynamic connectedness of biological and physical worlds are making natural risks increasingly systemic. Since risk is the result of individuals and collective decisions, the United Nations warn that, as risk gets more and more complex, responsibilities cannot be clearly assigned to the different stakeholders. Governments are able to influence risk-generating or risk-reducing behaviour in the population, in the private, public and voluntary sectors through public policies, hence \cite{GAR2019} argues that ``\textit{by incentivizing transdisciplinary, integrated, multisectoral research engaging non-traditional counterparts, risk assessment and decision-making efficiency can be improved, duplication of effort reduced, and connected collective action facilitated. National planning bodies with representation from all sectors must develop risk reduction strategies that assume an ``all-of-State'' institutions approach to risk reduction}''.

The United Nations encourage countries to establish solid partnerships both with private stakeholders and between governments, but, though necessary, they can also be difficult to manage. In particular, a United Nations survey on public-private partnerships for natural risk management \cite{UNISDR2} revealed that an active participation of all the community at risk is fundamental for the effectiveness of a strategy, but it can be achieved only if the government is able to create the proper conditions. First of all, arousing the interest of the private sector is necessary, as firms will participate in the initiatives only if they deem them convenient, and maintain their commitment over time. Since a plurality of private subjects make up the community, disaster risk management should be mutually beneficial and local authorities must encourage active and productive dialogue among the subjects involved (for example, between insurance and academia). In order to involve private companies, the government can ask them to develop a project and make them responsible for the parts that compete with them. Furthermore, the role of individual citizens, who can make an important contribution if properly informed and involved, should not be overlooked. In fact, educating the right segment of the population can encourage risk reduction behaviors, and fosters the dialogue between individuals and authorities. It is also important that citizens believe in the risk management strategy, so that they will actively engage in safety and involve more and more people. Trust between institutions and partners is essential for the success of the strategy and it is the duty of the government to constantly nourish it (for example by punishing the fraudulent companies that offer individuals risk reduction goods or services at excessively high prices), but constant monitoring private partners is also crucial.

A ``community-based disaster risk management"\footnote{ \cite{WHO} defines a community as ``\textit{a group of people living in the same environment, sharing the same livelihood. [...]. Community members are the immediate victims of the adverse effects of disasters and they have the best knowledge about their local surroundings such as demographic, social, economic and cultural status, risky areas, water sources, roads and health facilities. In addition community members have information about the vulnerable groups [...] and can assist health care}". Community-based disaster risk management empowers the community, address the root causes of risks and address it through local knowledge and expertise and for this reason activities and actions vary from one community to another \citep{Heijmans}.} raises the voice of people that are hence able to better communicate vulnerabilities to the governor, thus allowing for a more comprehensive view of risk. Understanding the social context is in fact essential for the construction of effective measures, and it is particularly important to reach out the weakest segments of the population.\footnote{
Some studies have shown that the poorests are more vulnerable to natural disasters due to their limited ability to cope with disasters  \citep{WB2}. As the impact of natural disaster on well-being might also be tremendous, other factors impacting vulnerability are inequalities concerning gender, age, education, ethnicity, wealth, health status, disability, access to resources and environmental concerns \citep{WB4, IPCC2}.
} In particular, \cite{Twigg} claims that ``\textit{participatory risk reduction initiatives are more sustainable because they build on local capacity, ideas can be tested and refined before adoption, and they are more likely to be compatible with long-term development plans. They may also be more cost-effective in the long term than externally-driven initiatives}". Unfortunately, in \cite{Twigg2}, the author argues that reality diverges from the desirable scenario of trust, collaboration and dialogue. 

In addition, developing countries could enormously benefit from collaborations between national governments and international institutions such as the World Bank, which for their part are interested in promoting risk reduction to contain future expenses \citep{KunreutherWB}.
Most regions with high exposure to natural hazards are involved in projects with intergovernmental organizations coordinating the disaster risk management \citep{GAR2019}, though some advancement is still needed. On the negative side, sometimes a lack of trust has hindered collaborative preparedness effort in conjunction with international aid agencies \citep{Twigg}, and collaboration between national governments should be strengthened.


\section{Final Remarks and Future Challenghes}
\label{Future}

In 2015, the United Nations declared that natural hazards could erase decades of progress \citep{UNAssembly}, but development itself fosters natural hazards. In fact, unsustainable development leads to an increase in social inequalities and in the number of poor who, because of their limited resources, constitute the most vulnerable and least resilient segment of the population. In turn, inequalities create social and political exclusion, and since the participation of the whole community is necessary for risk understanding, this poses a further obstacle to the effectiveness of the managers' choices \citep{GAR2019}.

Especially in the poorest countries, development went along with rapid urbanization, which has strong impacts on risks. Cities are in fact extremely dangerous due to biological, chemical, physical and socio-political conditions. In addition, they host large numbers of low-income people who live in low-quality facilities, sometimes without access to adequate infrastructure. The urbanization process, often carried out in an uncontrolled way, has therefore brought together a large number of extremely vulnerable people in highly risky areas \citep{WB3}. In this respect, \cite{WB2} registered a global trend toward increased risk taking: ``\textit{from 1970 to 2010 the world population grew by 87 percent, while the population in flood plains increased by 114 percent and in cyclone-prone coastlines by 192 percent}". The situation aggravates when appropriate building codes are lacking. 

Further consequences of urbanization are damage to the ecosystem and deforestation, which can cause the emergence of new natural phenomena, such as the spread of new diseases \citep{Opitz-Stapleton}. Climate change runs faster every year and its consequences are more and more tangible all over the world. According to \cite{GenevaAssociation,GenevaAssociation2}, the changing climate is challenging the weather-relater branches of the insurance industry. State-owned insurance and reinsurance companies are particularly fragile since, being bound by subsidy rules and government choices, are less flexible than private ones and will not be able to adapt their prices quickly \citep{Penning-Rowsell,Olcina,van_Oldenborgh_2017}. According to \cite{WB5}, climate change could drag up to 100 million people into poverty by 2030, but the financing needs for adaptation in developing countries far exceed funds available \citep{WB12}. Country's reluctance in investing in adaptation is largely due to the uncertainty surrounding the projects' benefits and, in addition, government undertaking some project may not be well-rewarded by the population, whose support or opposition is determined by the perception of danger associated to climate change or to the use of current technologies \citep{KunreutherGupta}. A first step that governments can take toward building a more climate-resilient society is encouraging or forcing private initiatives through adequate spatial planning frameworks, infrastructure projects and policy appraisals, regulatory and economic standards \citep{OECD5}. Public-private partnerships can channel and coordinate the efforts and objectives of governments and multiple private entities - project companies, lenders, shareholders, insurers and professional advisors - provided that tasks and responsibilities are clearly assigned \citep{WB13}. In order for the plan to be effective, governments should ensure consistency with the other policies adopted which could otherwise distort the climate-resilient incentives and discourage the adoption of the desired solutions \citep{OECD5}. As argued by the \cite{WB5}, we hence need to rapidly switch to inclusive climate-informed development. 

In light of this, it is increasingly evident that the management of natural risks cannot be effectively achieved unless included in a sustainable development plan. \cite{GAR2019} warns that ``\textit{with increasing complexity and interaction of human, economic and political systems within ecological systems, risk becomes increasingly systemic. [...]. The way in which such changes – including in the intensity and frequency of hazards – affect human activity is as yet difficult to foresee. Current approaches to risk measurement and management are inadequate to meet the challenges of the multifaceted interconnectedness of hazard, the barely understood breadth of exposure, and the profound detail of vulnerability}''.

A ``development-enhancing risk-management'' includes risk reduction in the development plans, allowing to define a set of coherent actions and thus avoiding those conflicts of interest that have led to a scarce commitment of local authorities in risk mitigation. Furthermore, it takes into account all the externalities of a risk reduction measure, some of which may not concern natural risks \citep{Clarke,Kunreuther2006,OECD5}. A first step in embracing this comprehensive approach might be incorporating risk reduction measures into existing funding streams \citep{Twigg}, for example by activating a coupled loan and voucher program for homeowners to relocate out of risk-prone areas \citep{KouskyKunreuther}. Governments might also prevent risk creation and foster mitigation by setting appropriate development-related standards and regulations, policies on social protection and payment for ecosystem services.
Among financial tools, in addition to traditional reinsurance and ILS, state-contingent debt instruments can help dealing with the post-disaster phase, while green bonds can foster climate change adaptation \citep{Opitz-Stapleton}.


\bibliography{main}

@article{Charpentier,
title = "Natural catastrophe insurance: How should the government intervene?",
journal = "Journal of Public Economics",
volume = "115",
pages = "1 - 17",
year = "2014",
issn = "0047-2727",
doi = "https://doi.org/10.1016/j.jpubeco.2014.03.004",
url = "http://www.sciencedirect.com/science/article/pii/S004727271400053X",
author = "Charpentier, Arthur and {Le Maux}, Benoit, "
}

@book{Consorcio,
  author       = {{Consorcio de Compensación de Seguros}}, 
  title        = {Natural Catastrophes Insurance Cover. A Diversity of Systems},
  publisher    = {Consorcio de Compensación de Seguros},
  year         = {2008}
}

@book{Consorcio2,
  author       = {{Consorcio de Compensación de Seguros}}, 
  title        = {La cobertura de los riesgos extraordinarios en España},
  publisher    = {Consorcio de Compensación de Seguros},
  year         = {2017}
}

@incollection{Grossi,
  author={Grossi, Patricia and Kunreuther, Howard and Windeler, Don},
  year= 2005, 
  booktitle={An Introduction to Catastrophe Models and Insurance}, 
  editor = {Grossi, Patricia and Kunreuther, Howard}, 
  title= {Catastrophe Modeling: A New Approach to Managing Risk}, 
  publisher= {Springer}
}

@article{Kousky,
author="Kousky, Carolyn and Cooke, Roger",
title="Explaining the Failure to Insure Catastrophic Risks",
journal="The Geneva Papers on Risk and Insurance - Issues and Practice",
year="2012",
month="Apr",
day="01",
volume="37",
number="2",
pages="206-227",
issn="1468-0440",
doi="10.1057/gpp.2012.14",
url="https://doi.org/10.1057/gpp.2012.14"
}

@article{Kousky2,
author="Kousky, Carolyn and Cooke, Roger",
title="The unholy Trinity: fat tails, tail dependence, and micro-correlations",
journal="Resources for the Future, Washington, RFF DP 09-36-REV",
year="2009",
month="11",
day="09"
}

@article{Kunreuther,
author = {Kunreuther, Howard},
year = {1996},
volume = {12 (2/3)},
pages = {171-187},
title = {Mitigating Disaster Losses through Insurance},
journal = {Journal of Risk and Uncertainty, Special Issue: The Stanford University Conference on Social Treatment of Catastrophic Risk
}
}

@article{Kunreuther2,
author = {Kunreuther, Howard},
year = {2006},
volume = {614},
pages = {208-227},
title = {Disaster mitigation and insurance: Learning from Katrina.},
journal = {Annals of American Academy of Political and Social Science
}
}

@article{Kunreuther3,
author = {Kunreuther, Howard},
year = {2015},
volume = {40},
pages = {741–762},
title = {The Role of Insurance in Reducing Losses from Extreme Events: The Need for Public–Private Partnerships.},
journal = {The Geneva Papers on Risk and Insurance - Issues and Practice
}
}

@article{Meyers,
title = "The competitive market equilibrium risk load formula for catastrophe ratemaking",
journal = "Proceedings of the Casualty Actuarial Society",
volume = "LXXXIII",
pages = "563-600",
year = "1996",
author = "G.G. Meyers"
}

@book{WB,
  author       = {{World Bank}}, 
  title        = {Financial Protection against Natural Disaster: An Operational Framework for Disaster Risk Financing and Insurance scholar},
  publisher    = {World Bank Group},
  year         = {2014}
}

@book{GenevaAssociation,
  author       = {{Geneva Association}}, 
  title        = {Warming of the Oceans and Implications for the (Re)insurance Industry. A Geneva Association Report},
  publisher    = { Geneva: Geneva Association},
  year         = {2013}
}

@book{GenevaAssociation2,
  author       = {{Geneva Association}}, 
  title        = {Climate Change and the Insurance Industry: Taking Action as Risk Managers and Investors. Perspectives from C-level executives in the insurance industry.},
  publisher    = {Geneva: Geneva Association},
  month = {1},
  year         = {2018}
}

@article{Feldblum,
title = "Risk loads for insurers",
journal = "Proceedings of the Casualty Actuarial Society",
volume = "LXXVII",
pages = "160-195",
year = "1990",
author = "S. Feldblum"
}

@article{Mango97,
title = "An application of game theory: property catastrophe risk load",
journal = "Proceedings of the Casualty Actuarial Society",
volume = "Spring 1997",
pages = "33-49",
year = "1997",
author = "D. Mango"
}

@article{Mango98,
title = "The concentration charge: reflecting catastrophe exposure accumulation in rates",
journal = "Proceedings of the Casualty Actuarial Society",
volume = "Winter 1998",
pages = "195-207",
year = "1998",
author = "D. Mango"
}

@article{Kreps,
title = "Reinsurer risk loads from marginal surplus requirements",
journal = "Proceedings of the Casualty Actuarial Society",
volume = "LXXVII",
pages = "196-203",
year = "1990",
author = "R. Kreps"
}

@article{Kreps2,
title = "Investment-Equivalent Reinsurance Pricing",
journal = "Proceedings of the Casualty Actuarial Society",
month = "5",
year = "1998",
author = "R. Kreps"
}

@book{Mitchell-Wallace,
  author       = {K. Mitchell-Wallace and M. Foote and J. Hillier and M. Jones}, 
  title        = {"Natural catastrophe risk management and modelling: A practitioner's guide"},
  publisher    = {Wiley Blackwell},
  year         = {2017}
}

@book{OECD2,
  author       = "OECD", 
  title        = "Disaster Risk Financing: A global survey of practices and challenges",
  publisher    = "OECD Publishing, Paris",
  year         = "2015",
  url           = "http://dx.doi.org/10.1787/9789264234246-en"
}

@book{OECD1,
  author       = "OECD", 
  title        = "Disaster Risk Assessment and Risk Financing. A G20/OECD Methodological Framework",
  publisher    = "G20 meeting in Mexico City",
  day = "4-5",
  month = "11",
  year         = "2012"
}

@article{UNISDR,
    author = {UNISDR},
    title = "{National Disaster Risk Assessment - Words into Action Guidelines - Governance System, Methodologies, and Use of Results}",
    journal = "The United Nations Office for Disaster Risk Reduction, Geneva, Switzerland",
    year = {2017}
}

@book{Sendai,
  author       = "UNISDR", 
  title        = "Sendai Framework for Disaster Risk Reduction 2015 - 2030",
  publisher    = "The United Nations Office for Disaster Risk Reduction, Geneva, Switzerland",
  year         = "2015"
}

@article{Mechler,
author = {Mechler, Reinhard},
year = {2016},
month = {02},
pages = {},
title = {Reviewing estimates of the economic efficiency of disaster risk management: opportunities and limitations of using risk-based cost–benefit analysis},
volume = {81},
journal = {Natural Hazards},
doi = {10.1007/s11069-016-2170-y}
}

@inbook{IPCC1,
author = {Field, C.B. and V. Barros and T.F. Stocker and D. Qin, D.J. Dokken and K.L. Ebi, M.D. Mastrandrea and K.J. Mach and G.-K. Plattner and S.K. Allen and M. Tignor, and P.M. Midgley},
year = {2012},
month = {6},
pages = {},
title = {Managing the Risks of Extreme Events and Disasters to Advance Climate Change Adaptation.},
publisher = {Cambridge University Press}
}

@article{Glauber,
author = {Glauber, Joseph},
year = {2004},
month = {02},
pages = {1179-1195},
title = {Crop Insurance Reconsidered},
volume = {86},
journal = {American Journal of Agricultural Economics},
doi = {10.1111/j.0002-9092.2004.00663.x}
}

@article{Hyogo,
author = "{United Nations}",
year = {2005},
title = {Hyogo Framework for Action 2005-2015: Building the Resilience of Nations and Communities to Disasters},
journal = {Extract from the final report of the World Conference on Disaster Reduction (A/CONF.206/6)}
}

@article{Matthews,
author = "Matthews, P. B. and Sheffield, M. and Andre, J. and Lafayette, J. and Roethen, J. and Dobkin, E.",
year = {1999},
title = {Insolvency: Will Historic Trends Return?},
journal = {Best’s Review-Property-Casualty Insurance Edition}
}

@article{Mills,
author = "Mills, E. and Lecomte, E. and Peara., A.",
year = {2001},
title = {U.S. Insurance Industry Perspectives on Global Climate Change.},
journal = {Best’s Review-Property-Casualty Insurance Edition}
}

@Article{Froot,
  author={Froot, Kenneth A.},
  title={{The market for catastrophe risk: a clinical examination}},
  journal={Journal of Financial Economics},
  year=2001,
  volume={60},
  number={2-3},
  pages={529-571},
  month={May},
  doi={},
  url={https://ideas.repec.org/a/eee/jfinec/v60y2001i2-3p529-571.html}
}

@article{McAneney,
title = "Government-sponsored natural disaster insurance pools: A view from down-under",
journal = "International Journal of Disaster Risk Reduction",
volume = "15",
pages = "1 - 9",
year = "2016",
issn = "2212-4209",
doi = "https://doi.org/10.1016/j.ijdrr.2015.11.004",
url = "http://www.sciencedirect.com/science/article/pii/S221242091530159X",
author = "John McAneney and Delphine McAneney and Rade Musulin and George Walker and Ryan Crompton"
}

@article{Jaffee,
title = "The Welfare Economics of Catastrophe Losses and Insurance",
journal = "The Geneva Papers on Risk and Insurance - Issues Practice",
volume = "38",
pages = "469-494",
year = "2013",
doi = "https://doi.org/10.1057/gpp.2013.17",
author = "Jaffee, D. and Russell, T."
}

@article{Burby,
author = {Raymond J. Burby},
year = {2001},
volume = {3},
issue = {3-4},
month = {09-12},
pages = {111-122},
title = {Behavioral models of insurance: The case of the California Earthquake Authority},
journal = {Flood insurance and floodplain management: the US experience}
}

@article{USAccounting,
author = "{US General Accounting Office}",
year = {2001},
month = {07},
day = {19},
pages = {2-3},
title = {Flood Insurance, Information on the Financial Condition of the National Flood Insurance Programme}
}

@book{Hallegatte,
author = { Hallegatte, Stéphane },
title = {How economic growth and rational decisions can make disaster losses grow faster than wealth},
publisher = {The World Bank},
year = {2011},
doi = {10.1596/1813-9450-5617},
address = {},
edition   = {},
URL = {https://elibrary.worldbank.org/doi/abs/10.1596/1813-9450-5617},
eprint = {https://elibrary.worldbank.org/doi/pdf/10.1596/1813-9450-5617}
}

@article{Cummins1,
author = {Cummins, J.D.},
year = {2007},
volume = {10},
issue = {2},
pages = {179–220},
title = {Reinsurance for Natural and Man-Made Catastrophes in the United States: Current State of the Market and Regulatory Reforms},
journal = {Risk Management and Insurance Review}
}

@article{Cummins2,
author = {Cummins, J.D.},
year = {2006},
volume = {88},
issue = {4},
pages = {337–79},
title = {Should the Government Provide Insurance for Catastrophes?},
journal = {Federal Reserve Bank of St. Louis Review}
}

@article{Cummins3,
author = {Cummins, J.D.},
year = {2008},
volume = {11},
pages = {23-47},
title = {Cat bonds and other risk-linked securities: state of the market and recent developments},
journal = {Risk Management and Insurance Review}
}

@incollection{CumminsBarrieu,
  author={Cummins, J.D. and Barrieu, P.},
  year= 2012, 
  title={Innovations in insurance markets: Hybrid and securitized risk-transfer solutions},
  editor = {George Dionne (Ed.)}, 
  booktitle= {Handbook of Insurance (2nd ed.)},
  publisher= {Kluwer Academic Publishers, Boston}
}

@article{LeeYu,
author = {Lee, Jin-Ping and Yu, Min-Teh},
year = {2007},
month = {09},
pages = {264-278},
title = {Valuation of Catastrophe Reinsurance With Catastrophe Bonds},
volume = {41},
journal = {Insurance: Mathematics and Economics},
doi = {10.1016/j.insmatheco.2006.11.003}
}

@book{GAR2019,
author = "{UNDRR}",
title = {Global Assessment Report on Disaster Risk Reduction 2019},
publisher = {United Nations, Geneva, Switzerland, United Nations Office for Disaster Risk Reduction},
year = {2019}
}

@article{Opitz-Stapleton,
author = {Opitz-Stapleton, Sarah and Nadin, Rebecca and Kellett, Jan and Calderone, Margherita and Quevedo, Adriana and Peters, Katie and Mayhew, Leigh},
year = {2019},
month = {05},
pages = {},
title = {Risk-informed development. From crisis to resilience},
journal = {Overseas Development Institute},
doi = {10.13140/RG.2.2.24990.28480}
}

@article{Rozenberg,
author = {Julie Rozenberg and Marianne Fay},
year = {2019},
title = {Beyond the Gap : How Countries Can Afford the Infrastructure They Need while Protecting the Planet. Sustainable Infrastructure},
journal = {Washington, DC: World Bank. © World Bank},
URL = {https://openknowledge.worldbank.org/handle/10986/31291}
}

@article{Twigg,
author = {Twigg, J.},
year = {2015},
title = {Disaster Risk Reduction. Good Practice Review 9},
journal = {Overseas Development Institute},
URL = {https://goodpracticereview.org/wp-content/uploads/2015/10/GPR-9-web-string-1.pdf}
}

@article{Twigg2,
author = {Twigg, J.},
year = {2011},
title = {Physician, Heal Thyself? The Politics of Disaster Mitigation},
journal = {London: UCL Hazard Centre},
URL = {http://www.ucl.ac.uk/hazardcentre/resources/working_papers/working_papers_folder/wp1}
}

@article{Wilkinson2,
author = {Wilkinson, E. and J. Twigg and L. Weingärtner and K. Peters},
year = {2017},
title = {Delivering Disaster Risk Reduction by 2030:
Pathways to Progress},
journal = {Overseas Development Institute},
URL = {https://www.odi.org/sites/odi.org.uk/files/
resource-documents/11532.pdf}
}

@article{FONDEN,
author = "{World Bank}",
year = {2012},
title = {FONDEN : Mexico's Natural Disaster Fund - A Review},
journal = {World Bank, Washington, DC. © World Bank. License: CC BY 3.0 IGO.},
URL = {https://openknowledge.worldbank.org/handle/10986/26881}
}

@book{UNISDR2,
author = "UNISDR",
title = {Private Sector Activities in DRR: Good Practices and Lessons Learned},
publisher = {UNISDR, Bonn: UNISDR},
year = {2008},
URL = {http://www.unisdr.org/we/inform/publications/7519}
}

@article{Heijmans,
author = {Heijmans, Annelies},
year = {2009},
month = {01},
pages = {},
title = {The Social Life of Community-Based Disaster Risk Reduction: Origins, Politics and Framing, Disaster Studies},
journal = {Disaster Studies Working Paper 20. London: Aon Benfield UCL Hazard Research Center}
}

@book{WHO,
author = "{World Health Organization}",
year = {2015},
title = {Managing disaster risks in communities: a community-based approach to disaster risk reduction: training manual for the trainers of cluster representatives and volunteers.},
publisher = {World Health Organization. Regional Office for the Eastern Mediterranean},
URL = {https://apps.who.int/iris/handle/10665/204677}
}

@article{Cavallo,
author = {Cavallo, Eduardo and Noy, Ilan},
year = {2009},
month = {12},
pages = {},
title = {The Economics of Natural Disasters - A Survey},
volume = {5},
journal = {University of Hawaii at Manoa, Department of Economics, Working Papers},
doi = {10.2139/ssrn.1817217}
}

@article{Achen,
author = {Achen, Christopher and Bartels, Larry},
year = {2004},
month = {01},
title = {Blind Retrospection: Electoral Responses To Drought, Flu, and Shark Attacks},
journal = {Princeton University. Working Paper}
}

@article{healy_malhotra_2009,
title={Myopic Voters and Natural Disaster Policy}, volume={103},
DOI={10.1017/S0003055409990104},
number={3},
journal={American Political Science Review}, publisher={Cambridge University Press},
author={Healy, Andrew and Malhotra, Neil},
year={2009},
pages={387–406}}

@article{Cole,
author = {Cole, Shawn and Healy, Andrew and Werker, Eric},
year = {2012},
month = {03},
title = {Do Voters Demand Responsive Governments? Evidence from Indian Disaster Relief},
volume = {97},
journal = {Journal of Development Economics},
doi = {10.1016/j.jdeveco.2011.05.005}
}

@article{UNAssembly,
author = "{United Nations General Assembly}",
year = {2015},
title = {A/RES/69/283},
url = { https://www.un.org/en/development/desa/population/migration/generalassembly/docs/globalcompact/A_RES_69_283.pdf}
}

@book{WB3,
author = {Dickson, Eric; Baker, Judy L.; Hoornweg, Daniel; Tiwari, Asmita},
year = {2012},
title = {Urban Risk Assessments : Understanding Disaster and Climate Risk in Cities (English). Urban development series},
publisher = {Washington, DC: World Bank},
URL = {http://documents.worldbank.org/curated/en/659161468182066104/Urban-risk-assessments-understanding-disaster-and-climate-risk-in-cities}
}

@article{OECD3,
author = "OECD",
year = {2010},
title = {Catastrophe-Linked Securities and Capital Markets},
journal = {OECD High Level Advisory Board on Financial Management of Large-scale Catastrophes. Paris},
}

@article{Michel-Kerjan,
author = {Michel-Kerjan, Erwann and I. Zelenko and V. Cardenas and D. Turge},
year = {2011},
title = {Catastrophe Financing for Governments: Learning from the 2009-2012 MultiCat Program in Mexico},
journal = {OECD Working Papers on Finance,
Insurance and Private Pensions, No. 9, OECD Publishing},
doi = {10.1787/5kgcjf7wkvhb-en}
}

@book{Michel-Kerjan2,
author = {Michel-Kerjan, Erwann and P. Slovic},
year = {2010},
title = {The Irrational Economist},
publisher = {New York: Public Affairs Press.}
}

@book{WB2,
author = {Stéphane Hallegatte and Adrien Vogt-Schilb and Mook Bangalore and Julie Rozenberg},
year = {2017},
title = {Unbreakable: Building the Resilience of the Poor in the Face of Natural Disasters},
publisher = {The World Bank}
}

@book{WB4,
author = {Hallegatte, Stéphane and Rentschler, Jun and Walsh, Brian},
year = {2018},
title = {Building Back Better : Achieving Resilience through Stronger, Faster, and More Inclusive Post-Disaster Reconstruction},
publisher = {The World Bank, Washington, DC}
}

@book{WB5,
author = {Stéphane Hallegatte and Mook Bangalore and Laura Bonzanigo and Marianne Fay and Tamaro Kane and Ulf Narloch and Julie Rozenberg and David Treguer and Adrien Vogt-Schilb},
year = {2016},
title = {Shock Waves Managing the Impacts of Climate Change on Poverty},
publisher = {The World Bank, Washington, DC}
}

@book{IPCC2,
author = {P.R. Shukla and J. Skea and E. Calvo Buendia and V. Masson-Delmotte and H.-O. Pörtner and D. C. Roberts and P. Zhai and R. Slade and S. Connors and R. van Diemen and M. Ferrat and E. Haughey and S. Luz and S. Neogi and M. Pathak and J. Petzold and J. Portugal Pereira and P. Vyas and E. Huntley and K. Kissick and M. Belkacemi and J. Malley},
year = {2019},
title = {Climate Change and Land: an IPCC special report on climate change, desertification, land degradation, sustainable land
management, food security, and greenhouse gas fluxes in terrestrial ecosystems},
publisher = {IPCC}
}

@article{Woodard,
author = {Woodard, Joshua and Schnitkey, Garu and Sherrick, Bruce and Lozano-Gracia, Nancy and Anselin, Luc},
year = {2012},
month = {03},
pages = {},
title = {A Spatial Econometric Analysis of Loss Experience in the U.S. Crop Insurance Program},
volume = {79},
journal = {Journal of Risk \& Insurance},
doi = {10.1111/j.1539-6975.2010.01397.x}
}

@article{KunreutherLyster,
author = {Kunreuther, Howard and Lyster, Rosemary},
year = {2016},
month = {09},
pages = {29-54},
title = {The role of public and private insurance in reducing losses from extreme weather events and disasters},
volume = {19},
journal = {Asia Pacific Journal of Environmental Law},
doi = {10.4337/apjel.2016.01.02}
}

@incollection{Linnerooth-Bayer,
  author      = "Linnerooth-Bayer, Joanne and Surminski, S. and Bouwer, L.M. and Noy, I. and Mechler, R.",
  title       = "Insurance as a Response to Loss and Damage?",
  editor      = "Mechler, R. and Bouwer, L. and Schinko, T. and Surminski, S. and Linnerooth-Bayer, J.",
  booktitle   = "Loss and Damage from Climate Change. Climate Risk Management, Policy and Governance",
  publisher   = "Springer",
  year        = 2019
}

@article{Kousky2018,
title = "Does federal disaster assistance crowd out flood insurance?",
journal = "Journal of Environmental Economics and Management",
volume = "87",
pages = "150 - 164",
year = "2018",
issn = "0095-0696",
doi = "https://doi.org/10.1016/j.jeem.2017.05.010",
url = "http://www.sciencedirect.com/science/article/pii/S0095069617303479",
author = "Carolyn Kousky and Erwann O. Michel-Kerjan and Paul A. Raschky"
}

@article{Hudson,
title = "Incentivising flood risk adaptation through risk based insurance premiums: Trade-offs between affordability and risk reduction",
journal = "Ecological Economics",
volume = "125",
pages = "1 - 13",
year = "2016",
issn = "0921-8009",
doi = "https://doi.org/10.1016/j.ecolecon.2016.01.015",
url = "http://www.sciencedirect.com/science/article/pii/S0921800916301240",
author = "Paul Hudson and W.J. Wouter Botzen and Luc Feyen and Jeroen C.J.H. Aerts"
}

@article{Gan,
title = "Wildfire risk adaptation: Propensity of forestland owners to purchase wildfire insurance in the southern United States",
journal = "Canadian Journal of Forest Research",
volume = "44(11)",
pages = "1376–1382",
year = "2014",
doi = "https://doi.org/10.1139/cjfr-2014-0301",
author = "Gan, J. and A. Jarrett and C.J. Gaither"
}

@article{McIntosh,
title = "Productivity, credit, risk, and the demand for weather index insurance in smallholder agriculture in Ethiopia",
journal = "Agricultural Economics",
volume = "44",
pages = "399-417",
year = "2013",
doi = "doi:10.1111/agec.12024",
author = "McIntosh, C. and Sarris, A. and Papadopoulos, F."
}

@article{Greatrex,
title = "Scaling up index insurance for smallholder farmers: Recent evidence and insights",
journal = "CCAFS Report No. 14. Copenhagen, Denmark: CGIAR Research Program on Climate Change, Agriculture and Food Security (CCAFS)",
year = "2015",
author = "Greatrex, H. and Hansen, J. and Garvin, S. and Diro, R. and Blakeley, S. and Le Guen, M. and Rao, K. and Osgood, D."
}

@article{Conradt,
title = "Flexible weather index-based insurance design",
journal = "Climate Risk Management",
volume = "10",
pages = "106 - 117",
year = "2015",
issn = "2212-0963",
doi = "https://doi.org/10.1016/j.crm.2015.06.003",
url = "http://www.sciencedirect.com/science/article/pii/S2212096315000212",
author = "Sarah Conradt and Robert Finger and Martina Spörri"
}

@article{Bogale,
author = {Ayalneh Bogale},
title = {Weather-indexed insurance: an elusive or achievable adaptation strategy to climate variability and change for smallholder farmers in Ethiopia},
journal = {Climate and Development},
volume = {7},
number = {3},
pages = {246-256},
year  = {2015},
publisher = {Taylor & Francis},
doi = {10.1080/17565529.2014.934769},

URL = { 
        https://doi.org/10.1080/17565529.2014.934769
    
},
eprint = { 
        https://doi.org/10.1080/17565529.2014.934769
    
}

}

@article{Daron,
title = "Assessing pricing assumptions for weather index insurance in a changing climate",
journal = "Climate Risk Management",
volume = "1",
pages = "76 - 91",
year = "2014",
issn = "2212-0963",
doi = "https://doi.org/10.1016/j.crm.2014.01.001",
url = "http://www.sciencedirect.com/science/article/pii/S2212096314000023",
author = "J.D. Daron and D.A. Stainforth"
}

@article{Eling,
title = "The Determinants of Microinsurance Demand",
journal = "Geneva Paper on Risk and Insurance - Issues and Practice",
volume = "39",
pages = "224–263",
year = "2014",
doi = "https://doi.org/10.1057/gpp.2014.5",
author = "Eling, M. and Pradhan, S. and Schmit, J."
}

@incollection{Lotze-Campen,
author = {Lotze-Campen, Hermann and Popp, Alexander},
year = {2012},
month = {06},
pages = {171-178},
title = {Agricultural Adaptation Options: Production Technology, Insurance, Trade},
editor = {Edenhofer, O., J. Wallacher, H. Lotze-Campen, M. Reder, B. Knopf (eds.)},
booktitle   = {Climate Change, Justice and Sustainability},
publisher   = {Springer Netherlands, Dordrecht, Netherlands},
isbn = {978-94-007-4539-1},
doi = {10.1007/978-94-007-4540-7_16}
}

@article{Cole2,
title = "Overcoming Barriers to Microinsurance Adoption: Evidence from the Field",
journal = "Geneva Paper on Risk and Insurance - Issues and Practice",
volume = "40",
pages = "720–740",
year = "2015",
doi = "https://doi.org/10.1057/gpp.2015.12",
author = "Cole, Shawn"
}

@article{Cole3,
title = "Barriers to Household Risk Management: Evidence from India",
journal = "American Economic Journal: Applied Economics",
volume = "5 (1)",
pages = "104-35",
year = "2013",
author = "Cole, Shawn and Xavier Giné and Jeremy Tobacman and Petia Topalova and Robert Townsend and James Vickery"
}

@article{Gallagher,
 ISSN = {19457782, 19457790},
 URL = {http://www.jstor.org/stable/43189495},
 author = {Justin Gallagher},
 journal = {American Economic Journal: Applied Economics},
 number = {3},
 pages = {206--233},
 publisher = {American Economic Association},
 title = {Learning about an Infrequent Event: Evidence from Flood Insurance Take-Up in the United States},
 volume = {6},
 year = {2014}
}

@article{Siebert,
 author = {Siebert, A.},
 journal = {American Economic Journal: Applied Economics},
 pages = {15–28},
 publisher = {Climatic Change},
 title = {Analysis of the future potential of index insurance in the West African Sahel using CMIP5 GCM results},
 volume = {134},
 doi = "https://doi.org/10.1007/s10584-015-1508-x",
 year = {2016}
}

@article{Kleindorfer,
 author = {Kleindorfer, P.R. and Kunreuther, H. and Ou-Yang, C.},
 journal = {American Economic Journal: Applied Economics},
 pages = {51–78},
 publisher = {Journal of Risk and Uncertainty},
 title = {Single-year and multi-year insurance policies in a competitive market},
 volume = {45},
 doi = "https://doi.org/10.1007/s11166-012-9148-2",
 year = {2012}
}

@incollection{KunreutherGupta,
  author      = "Kunreuther, Howard and S. Gupta and V. Bosetti and R. Cooke and V. Dutt and M. Ha-Duong and H. Held and J. Llanes-Regueiro and A. Patt and E. Shittu and and E.
Weber",
  title       = "Integrated Risk and Uncertainty Assessment of Climate Change Response Policies",
  editor      = "Edenhofer, O. and R. Pichs-Madruga and Y. Sokona and E. Farahani and S. Kadner and K. Seyboth and A. Adler and I. Baum and S. Brunner and P. Eickemeier and B. Kriemann and J. Savolainen and S. Schlömer and C. von Stechow and T. Zwickel and J.C. Minx (eds.)",
  booktitle   = "Climate Change 2014: Mitigation of Climate Change. Contribution of Working Group III to the Fifth Assessment Report of the Intergovernmental Panel on Climate Change",
  publisher   = "Cambridge University Press, Cambridge, United Kingdom and New York, NY, USA",
  year        = 2014
}

@article{Dercon,
title = "Offering rainfall insurance to informal insurance groups: Evidence from a field experiment in Ethiopia",
journal = "Journal of Development Economics",
volume = "106",
pages = "132 - 143",
year = "2014",
issn = "0304-3878",
doi = "https://doi.org/10.1016/j.jdeveco.2013.09.006",
url = "http://www.sciencedirect.com/science/article/pii/S0304387813001338",
author = "Stefan Dercon and Ruth Vargas Hill and Daniel Clarke and Ingo Outes-Leon and Alemayehu Seyoum Taffesse"
}

@article{Cole4,
title = "The effectiveness of indexbased micro-insurance in helping smallholders manage weather-related risks",
journal = "London: EPPI-Centre, Social Science Research Unit, Institute of Education,University of London",
year = "2012",
author = "Cole, Shawn and Bastian, G. and Vyas, S. and Wendel, C. and Stein, D."
}

@article{Rao,
title = "Index based Crop Insurance",
journal = "Agriculture and Agricultural Science Procedia",
pages = "193-203",
year = "2010",
author = "Kolli N. Rao"
}

@book{Alderman,
author = {Alderman, Harold and Haque, Trina},
title = {Insurance Against Covariate Shocks},
publisher = {The World Bank},
year = {2007},
doi = {10.1596/978-0-8213-7036-0},
address = {},
edition   = {},
URL = {https://elibrary.worldbank.org/doi/abs/10.1596/978-0-8213-7036-0},
eprint = {https://elibrary.worldbank.org/doi/pdf/10.1596/978-0-8213-7036-0}
}

@article{Binswanger-Mkhize,
author = { Hans P.   Binswanger-Mkhize },
title = {Is There Too Much Hype about Index-based Agricultural Insurance?},
journal = {The Journal of Development Studies},
volume = {48},
number = {2},
pages = {187-200},
year  = {2012},
publisher = {Routledge},
doi = {10.1080/00220388.2011.625411},
URL = {https://doi.org/10.1080/00220388.2011.625411},
eprint = {https://doi.org/10.1080/00220388.2011.625411}
}

@book{WB6,
author = "{World Bank}",
title = {Managing Agricultural Production Risk: Innovations in Developing Countries},
publisher = {The World Bank - Agriculture and Rural Development Department},
year = {2005},
}

@article{KouskyKunreuther,
author = {Carolyn Kousky and Howard Kunreuther},
title = {Addressing Affordability in the National Flood Insurance Program},
journal = {Journal of Extreme Events},
volume = {01},
number = {1},
year  = {2014},
URL = {https://doi.org/10.1142/S2345737614500018}
}

@article{Michel-Kerjan3,
author = {Michel-Kerjan, Erwann and Forges, Sabine and Kunreuther, Howard},
year = {2011},
month = {09},
pages = {644-58},
title = {Policy tenure under the U.S. National Flood Insurance Program (NFIP)},
volume = {32},
journal = {Risk analysis : an official publication of the Society for Risk Analysis},
doi = {10.1111/j.1539-6924.2011.01671.x}
}

@article{Chivers,
author = {Chivers, James and Flores, Nicholas},
year = {2002},
month = {11},
pages = {},
title = {Market Failure in Information: The National Flood Insurance Program},
volume = {78},
journal = {Land Economics},
doi = {10.2307/3146850}
}

@article{Michel-Kerjan4,
author = {Michel-Kerjan, Erwann},
year = {2010},
month = {11},
pages = {165-86},
title = {Catastrophe Economics: The National Flood Insurance Program},
volume = {24},
journal = {Journal of Economic Perspectives},
doi = {10.2307/20799178}
}

@book{Dixon,
author = "Lloyd Dixon and Noreen Clancy and Seth A. Seabury and Adrian Overton",
title = {The National Flood Insurance Program’s Market Penetration Rate: Estimates and Policy Implications},
publisher = {RAND Infrastructure, Safety, and Environment and Institute for Civil Justice},
month = {02},
year = {2006},
}

@article{Kriesel,
author = {Kriesel, Warren and Landry, Craig},
year = {2004},
month = {02},
pages = {405-420},
title = {Participation in the National Flood Insurance Program: An Empirical Analysis for Coastal Properties},
volume = {71},
journal = {Journal of Risk and Insurance},
doi = {10.1111/j.0022-4367.2004.00096.x}
}

@article{Marshall,
author = {Daniel Marshall},
year = {2018},
month = {01},
volume = {21},
issue = {1},
pages = {73-116},
title = {An Overview of the California Earthquake Authority},
journal = {Risk Management and Insurance Review}
}

@article{Wood,
author = {Amy Wood and Ilan Noy and Miles Parker},
year = {2016},
month = {02},
volume = {79},
issue = {3},
title = {The Canterbury rebuild five years on from the Christchurch earthquake},
journal = {Reserve Bank of New Zealand Bulletin}
}

@article{Basbug,
author = {B. Burcak Başbuğ‐Erkan and Ozlem Yilmaz},
year = {2015},
month = {10},
volume = {39},
issue = {4},
pages = { 782-794},
title = {Successes and failures of compulsory risk mitigation: re‐evaluating the Turkish Catastrophe Insurance Pool},
journal = {Disasters}
}

@article{Cakti,
author = {Cakti, Eser and Erdik, M. and Sesetyan, Karin},
year = {2006},
month = {01},
pages = {},
title = {Expected Earthquake Losses to Buildings in Istanbul and Implications for the Performance of the Turkish Catastrophe Insurance Pool},
journal = {Proceedings of 2006 ECI Conference on Geohazards Lillehammer, Norway}
}

@incollection{Clarke,
  author      = "Caroline Clarke and Neil A. Doherty",
  title       = "Development-Enhancing Risk Management",
  editor      = "Eugene N. Gurenko",
  booktitle   = "Catastrophe Risk and Reinsurance: A Country Risk Management Perspective",
  publisher   = "Risk Books, Incisive Financial Publishing",
  year        = 2004
}

@incollection{Vallet,
  author      = "Suzanne Vallet",
  title       = "Insuring the Uninsurable: The French Natural Catastrophe Insurance System",
  editor      = "Eugene N. Gurenko",
  booktitle   = "Catastrophe Risk and Reinsurance: A Country Risk Management Perspective",
  publisher   = "Risk Books, Incisive Financial Publishing",
  year        = 2004
}

@incollection{vonLucius,
  author      = "Johann-Adrian {von Lucius}",
  title       = "A Reinsurer’s Perspective on the Turkish Catastrophe Insurance Pool (TCIP)",
  editor      = "Eugene N. Gurenko",
  booktitle   = "Catastrophe Risk and Reinsurance: A Country Risk Management Perspective",
  publisher   = "Risk Books, Incisive Financial Publishing",
  year        = 2004
}

@incollection{Takeda,
  author      = "Yuichi Takeda",
  title       = "Government as Reinsurers of Last Resort: The Japanese Experience",
  editor      = "Eugene N. Gurenko",
  booktitle   = "Catastrophe Risk and Reinsurance: A Country Risk Management Perspective",
  publisher   = "Risk Books, Incisive Financial Publishing",
  year        = 2004
}

@incollection{Seo,
  author      = "John Seo",
  title       = "Evidence of Market Response to Coverage Value in Some Major Catastrophe Insurance Programmes",
  editor      = "Eugene N. Gurenko",
  booktitle   = "Catastrophe Risk and Reinsurance: A Country Risk Management Perspective",
  publisher   = "Risk Books, Incisive Financial Publishing",
  year        = 2004
}

@incollection{Kunreuther2006,
author = {Kunreuther, Howard},
year = {2006},
title = {Has the time come for comprehensive natural disaster insurance?},
editor = {R.J. Daniel and D.F. Kettl and H. Kunreuther},
booktitle = {On Risk and Disaster},
publisher = {University of Pennsylvania Press, Philadelphia}
}

@article{Machetti,
author = {Ignacio Machetti},
year = {2004},
title = {The Spanish Experience in the Management of Extraordinary Risks, Including Terrorism},
journal = {Background Note of Conference on Catastrophic Risks and Insurance, 22-23 November 2004, OECD}
}

@article{Olcina,
author = {Olcina, J. and Sauri, D. and Hernández, M. and Ribas, A.},
year = {2016},
title = {Flood policy in Spain: a review for the period 1983-2013},
journal = {Disaster Prevention and Management: an International Journal},
pages = {41-58},
volume = {25/1}
}

@article{Linnerooth-Bayer2,
author = {Linnerooth-Bayer, Joanne and Mechler, Reinhard and Hochrainer-Stigler, Stefan},
year = {2011},
month = {01},
pages = {1:23},
title = {Insurance against Losses from Natural Disasters in Developing Countries. Evidence, Gaps and the Way Forward},
volume = {1},
journal = {Journal of Integrated Disaster Risk Management}
}

@article{Surminski,
author = {Swenja Surminski},
year = {2018},
title = {Fit for Purpose and Fit for the Future? An Evaluation of the UK's New Flood Reinsurance Pool},
volume = {21},
issue = {1},
journal = {Risk Management and Insurance Review},
}

@article{Surminski2,
author = {Surminski, Swenja and Eldridge, Jillian},
year = {2014},
month = {09},
pages = {},
title = {Flood insurance in England- An assessment of the current and newly proposed insurance scheme in the context of rising flood risk},
journal = {Journal of Flood Risk Management},
doi = {10.1111/jfr3.12127}
}

@article{Penning-Rowsell,
author = {Penning-Rowsell, Edmund},
year = {2015},
month = {08},
pages = {},
title = {Flood insurance in the UK: a critical perspective},
volume = {2},
journal = {Wiley Interdisciplinary Reviews: Water},
doi = {10.1002/wat2.1104}
}

@article{Bruggeman,
author = {Bruggeman, Véronique and Faure, Michael and Fiore, Karine},
year = {2010},
month = {07},
pages = {369-390},
title = {The Government as Reinsurer of Catastrophe Risks?},
volume = {35},
journal = {The Geneva Papers on Risk and Insurance - Issues and Practice},
doi = {10.1057/gpp.2010.10}
}

@article{KunreutherPauly3,
author = {Kunreuther, Howard and Pauly, Mark},
year = {2006},
pages = {101–116},
title = {Rules rather than discretion: Lessons from Hurricane Katrina},
volume = {33},
journal = {Journal of Risk and Uncertainty},
doi = {10.1007/s11166-006-0173-x}
}

@article{Yazici,
author = {Yazici, S.},
year = {2006},
title = {The turkish Catastrophe Insurance Pool TCIP and Compulsory Earthquake Insurance Scheme},
journal = {OECD, Catastrophic Risks and Insurance, OECD Publishing}
}

@book{WB8,
author = "{World Bank}",
year = {2011},
title = {Turkish catastrophe insurance pool : providing affordable earthquake risk insurance (English)},
publisher = {Disaster risk financing and insurance case study. Washington, D.C. : World Bank Group},
url = {http://documents.worldbank.org/curated/en/853431468188946296/Turkish-catastrophe-insurance-pool-providing-affordable-earthquake-risk-insurance}
}

@book{WB9,
author = "Eugene Gurenko and Rodney Lester and Olivier Mahul and Serap Oguz Gonugal",
year = {2006},
title = {Earthquake Insurance in Turkey: History of the Turkish Catastrophe Insurance Pool},
publisher = {World Bank Group}}

@book{WB10,
author = "{World Bank}",
year = {2019},
title = {Accelerating risk reduction through forwardlooking investments and policies in Romania},
publisher = {Results in Resilience Series, The World Bank Group}}

@book{WB11,
author = "{World Bank}",
year = {2012},
title = {Improving the assessment of disaster risks to strengthen financial resilience (English)},
publisher = {Washington, DC: World Bank}}

@book{NFIP,
author = "FEMA",
year = {2020},
volume = "9",
title = {The Watermark Financial Report, Fiscal Year 2020, First Quarter},
publisher = {FEMA}}

@article{van_Oldenborgh_2017,
	doi = {10.1088/1748-9326/aa9ef2},
	url = {https://doi.org/10.1088%2F1748-9326%2Faa9ef2},
	year = 2017,
	month = {dec},
	publisher = {{IOP} Publishing},
	volume = {12},
	number = {12},
	pages = {124009},
	author = {Geert Jan van Oldenborgh and Karin van der Wiel and Antonia Sebastian and Roop Singh and Julie Arrighi and Friederike Otto and Karsten Haustein and Sihan Li and Gabriel Vecchi and Heidi Cullen},
	title = {Attribution of extreme rainfall from Hurricane Harvey, August 2017},
	journal = {Environmental Research Letters}
}

@incollection{Kleindorfer2,
author = {Paul Kleindorfer and Patricia Grossi and Howard Kunreuther},
year = {2005},
title = {The Impact of Mitigation on Homeowners and Insurers: An Analysis of Model Cities},
editor = {Patricia Grossi and Howard Kunreuther},
booktitle = {An Introduction to Catastrophe Models and Insurance},
publisher = {Springer}
}

@book{OECD4,
author = "OECD",
year = {2018},
title = {The Contribution of Reinsurance Markets to Managing Catastrophe Risk},
publisher = {OECD},
url = {www.oecd.org/finance/the-contribution-of-reinsurance-markets-to-managing-catastrophe-risk.pdf}}

@article{Raschky,
author = {Raschky, Paul A. and Weck-Hannemann, Hannelore},
year = {2007},
month = {01},
pages = {321-329},
title = {Charity Hazard—A Real Hazard to Natural Disaster Insurance?},
volume = {7},
journal = {Environmental Hazards},
doi = {10.1016/j.envhaz.2007.09.002}
}

@article{Raschky2,
author = {Raschky, Paul A. and Reimund Schwarze and Manijeh Schwindt and Ferdinand Zahn},
year = {2013},
pages = {179-200},
title = {Uncertainty of Governmental Relief and the Crowding out of Flood Insurance},
volume = {54},
journal = {Environmental and Resource Economics volume},
url = {https://doi.org/10.1007/s10640-012-9586-y}
}

@incollection{KunreutherWB,
author = {Howard Kunreuther},
year = {2003},
title = {Interdependent disaster risks: the need for public-private partnerships},
editor = {Kreimer, Alcira; Arnold, Margaret; Carlin, Anne},
booktitle = {Building Safer Cities: The Future of Disaster Risk  (English). Disaster Risk Management series no. 3},
publisher = {Washington, DC: World Bank}
}

@book{WB12,
author = "{World Bank}",
year = {2010},
title = {Economics of adaptation to climate change : social synthesis report (English)},
publisher = {Washington, DC: World Bank}}

@book{WB13,
author = "{World Bank}",
year = {2016},
title = {Emerging Trends in Mainstreaming Climate Resilience in Large Scale, Multi-sector Infrastructure PPPs},
publisher = {International Bank for Reconstruction and Development/The World Bank}}

@book{OECD5,
author = "OECD",
year = {2018},
title = {Climate-resilient Infrastructure},
publisher = {OECD Environment Policy Paper No. 14},
url = {www.oecd.org/finance/the-contribution-of-reinsurance-markets-to-managing-catastrophe-risk.pdf}}

@book{WB14,
author = "{World Bank}",
year = {2013},
title = {Mexico MultiCat Bond : Transferring Catastrophe Risk to the Capital Markets. Disaster risk financing and insurance case study},
publisher = {Washington, DC. © World Bank},
url = {https://openknowledge.worldbank.org/handle/10986/22422}
}

\begin{landscape}

\begin{ThreePartTable}
\renewcommand{\TPTminimum}{0.7\linewidth}
\begin{TableNotes}
  \small
    \item[1] California Earthquake Authority, Audited Financial Statements 2018.
    \item[2] Government of Iceland, ACT 55/1992 on The Natural Catastrophe Insurance of Iceland after changes to NTI‘s legislation in July 2018.
    \item[3] New Zealand Government, “Earthquake Commission Act 1993”; Civil Defence – New Zealand, “Government Financial Support”, 2009.
    \item[4] \cite{Consorcio,Consorcio2,Machetti}.
    \item[5] Government of Taiwan, “Insurance Act”, art 138-1, 1999; Government  of Taiwan, “Enforcement Rules for Coinsurance and Risk Assumption Mechanism of Residential Earthquake Insurance”,  2001; Government of Taiwan, “Taiwan Residential Earthquake Insurance Fund Articles of Incorporation”, 2001; Government of Taiwan, “Regulations Governing Taiwan Residential Earthquake Insurance Fund”, 2001; Government of Taiwan, “Enforcement Rules for the Risk Spreading Mechanism of Residential Earthquake Insurance”, 2008; Taiwan Residential Earthquake Insurance Fund (TREIF), Annual Report 2015.
    \item[6] \cite{Yazici, WB8, WB9}; Turkish Government, Law no: 4452 “Measures to be taken Against Natural Disasters and Authorization in Regards to Arrangements to be made in Overcoming the Damage Caused by Natural Disasters”, 27/08/1999; Turkish Government, Decree Law no: 587 “Decree Law Relating to Compulsory Earthquake Insurance”, 27/12/1999; Turkish Government, Law no: 6305 “Catastrophe Insurance Law”, accepted 09/05/2012; Turkish Government, Tariff and instruction of compulsory earthquake insurance, Official Gazette 28512, 29 December 2012.
    \item[7] All-Hazard Authorities of the Federal Emergency Management Agency, “The National Flood Insurance Act of 1968” as amended 42 U.S.C 4001 et seq., sec 1366, Office of the General Counsel, August 1997; US Government, “Disaster Mitigation Act of 2000”, Public Law 106-390,30 October 2000; Federal Insurance and Mitigation Administration - FEMA, “FY 2016 Pre-Disaster Mitigation (PDM) Grant Program. Fact Sheet”, FEMA, 2016; United States Code, Title 42. The Public Health and Welfare, Chapter 68. Disaster Relief, “Robert T. Stafford Disaster Relief and Emergency Assistance Act”, Public Law 93-288, signed into law 23 November 1988, last amended April 2013.
\end{TableNotes}

\begin{longtable}{lllll}
\caption{State-owned insurance companies (examples).}\label{tab:State-owned Insurance}\\
  \toprule
Country, Name, Year & Peril & \begin{tabular}[c]{@{}l@{}}Position in the\\ Market\end{tabular} & Policy Details  & Risk Reduction                \\                   
\midrule
  \endfirsthead

\multicolumn{5}{c}{{\bfseries \tablename\ \thetable{} -- continued from previous
page}}\\
\toprule
Country, Name, Year & Peril & \begin{tabular}[c]{@{}l@{}}Position in the\\ Market\end{tabular} & Policy Details  & Risk Reduction                \\  
  \midrule    
\endhead

\midrule
\multicolumn{5}{r}{{Continued on next page}} \\
\midrule
\endfoot
\bottomrule
\insertTableNotes         
\endlastfoot     

\begin{tabular}[c]{@{}l@{}}California,\\ California Earthquake\\ Authority\tnote{1},\\ 1996\end{tabular}            & Earthquakes                                                                                                                                                & \begin{tabular}[c]{@{}l@{}}Co-insurance with\\ private insurers.\end{tabular}                                                     & \begin{tabular}[c]{@{}l@{}}California insurers are obliged to\\ offer earthquake policies, but they\\ can choose whether to co-operate\\ with CEA or not. Policy purchase\\ is voluntary.\\ Risk-based premium. Possibility to\\ choose deductibles and maximum\\ coverage.\end{tabular}                                                                                                                                           & \begin{tabular}[c]{@{}l@{}}Incentives for risk\\ mitigation\\ (discounted\\ premiums), \\ Reinsurance.\end{tabular}                                      \\ \hline
\begin{tabular}[c]{@{}l@{}}Iceland,\\ National Catastrophe\\ Insurance of Iceland\tnote{2},\\ 1975\end{tabular}     & \begin{tabular}[c]{@{}l@{}}Avalanches\\ Earthquakes\\ Floods\\ Landslides\\ Volcanic eruptions\end{tabular}                                                & \begin{tabular}[c]{@{}l@{}}Private insurers\\ collect and transfer\\ premiums, and\\ receive a\\ commission.\end{tabular}         & \begin{tabular}[c]{@{}l@{}}All buildings and contents insured\\ against fire are also insured against\\ catastrophe risks.\\ Fire insurance is compulsory.\\ Flat premium set by law.\\ 5\% Minimum deductible.\end{tabular}                                                                                                                                                                                                       & \begin{tabular}[c]{@{}l@{}}Reinsurance,\\ If the agency borrows\\ funds, such loans are\\ unconditionally\\ guaranteed by the\\ government.\end{tabular} \\ \hline
\begin{tabular}[c]{@{}l@{}}New Zealand,\\ Earthquake\\ Commission (EQC)\tnote{3},\\ 1993\end{tabular}               & \begin{tabular}[c]{@{}l@{}}Earthquakes\\ Floods\\ Hydrotermal activity\\ Natural landslides\\ Tsunamis\\ Volcanic eruptions\end{tabular}                   & \begin{tabular}[c]{@{}l@{}}Co-insurance with\\ private insurers.\end{tabular}                                                     & \begin{tabular}[c]{@{}l@{}}Specific building cathegory are\\ covered compulsorily and\\ automatically along with fire\\ insurance. If the building is not\\ insured for fire, is also not covered\\ for natural risks.\\ Flat rates. Applies fixed maxima\\ and deductibles.\end{tabular}                                                                                                                                          & \begin{tabular}[c]{@{}l@{}}Reinsurance,\\ Unlimited State\\ guarantee.\end{tabular}                                                                      \\ \hline
\begin{tabular}[c]{@{}l@{}}Spain,\\ Consorcio de\\ Compensacion de\\ Seguros\tnote{4},\\ 1954\end{tabular}          & \begin{tabular}[c]{@{}l@{}}Atypical cyclonic storms\\ Earthquakes\\ Extraordinary floods\\ Fall of meteorites\\ Tsunamis\\ Volcanic eruptions\end{tabular} & \begin{tabular}[c]{@{}l@{}}Private insurers\\ collect and transfer\\ premiums, and\\ receive a\\ commission.\end{tabular}         & \begin{tabular}[c]{@{}l@{}}Compulsorily included in personal\\ accident policies, life insurance\\ and some branches of property\\ damage.\\ Flat rates. Deductibles apply to\\ most for the covers.\end{tabular}                                                                                                                                                                                                                  & \begin{tabular}[c]{@{}l@{}}Unlimited State\\ guarantee.\end{tabular}                                                                                     \\ \hline
\begin{tabular}[c]{@{}l@{}}Taiwan,\\ Taiwan Residential\\ Earthquake\\ Insurance Fund\tnote{5},\\ 2001\end{tabular} & Earthquakes                                                                                                                                                & \begin{tabular}[c]{@{}l@{}}Co-insurance with\\ private insurers\end{tabular}                                                      & \begin{tabular}[c]{@{}l@{}}Compulsorily attached to all\\ residential fire insurance policies.\\ Flat rates. Applies maxima.\end{tabular}                                                                                                                                                                                                                                                                                          & \begin{tabular}[c]{@{}l@{}}Reinsurance,\\ State guarantee.\end{tabular}                                                                                  \\ \hline
\begin{tabular}[c]{@{}l@{}}Turkey,\\ Turkish Catastrophe\\ Insurance Pool\tnote{6},\\ 2000\end{tabular}             & Earthquakes                                                                                                                                                & \begin{tabular}[c]{@{}l@{}}Accredited\\ insurance companies\\ and agents arrange\\ policies on behalf\\ of the TCIP.\end{tabular} & \begin{tabular}[c]{@{}l@{}}Compulsory for certain types of\\ buildings and dwellings.\\ Partially risk based rates (5 risk\\ zone, 3 construction types).\\ Applies maxima.\end{tabular}                                                                                                                                                                                                                                           & \begin{tabular}[c]{@{}l@{}}Reinsurance,\\ Contingent Credit\\ Line.\end{tabular}                                                                         \\ \hline
\begin{tabular}[c]{@{}l@{}}U.S.,\\ National Flood\\ Insurance Program\tnote{7},\\ 1968\end{tabular}                 & Floods                                                                                                                                                     & \begin{tabular}[c]{@{}l@{}}Competes with\\ private insurers.\end{tabular}                                                         & \begin{tabular}[c]{@{}l@{}}Residential buildings and contents\\ policies are offered to those\\ communities that participate to the\\ program. Communities' can\\ participate to the program but are\\ not forced. If they do, they\\ undertake to adopt appropriate\\ preventive measures. Members of a\\ community involved are not forced\\ to buy policies.\\ Risk-based rates. Applies maxima\\ and deductibles.\end{tabular} & \begin{tabular}[c]{@{}l@{}}Incetives for risk\\ mitigation,\\ Government as\\ lender of last resort.\end{tabular}                                        \\ \hline

\end{longtable}
\end{ThreePartTable}

\begin{ThreePartTable}
\renewcommand{\TPTminimum}{0.7\linewidth}
\begin{TableNotes}
  \small
    \item[1] Florida Hurricane Catastrophe Fund, Annual Report of Aggregate Net Probable Maximum Losses, Financing Options, and Potential Assessments, February 2020.
    \item[2] Caisse Centrale de Réassurance, Activity Report 2018.
    \item[3] Japan Earthquake Reinsurance, Annual Report 2019, Introduction to Earthquake Reinsurance in Japan.
    \item[4] Flood Re, Annual Report and Financial Statements, Year ended 31 March 2019.
\end{TableNotes}

\begin{longtable}{lllll}
\caption{State-owned reinsurance companies (examples).}\label{tab:State-owned Reinsurance}\\
  \toprule
\begin{tabular}[c]{@{}l@{}}Country, Name,\\ Year\end{tabular} & Peril & Policy Details & Reinsurance Contract & Risk Reduction  \\                   
\midrule
  \endfirsthead

\multicolumn{5}{c}{{\bfseries \tablename\ \thetable{} -- continued from previous
page}}\\
\toprule
\begin{tabular}[c]{@{}l@{}}Country, Name,\\ Year\end{tabular} & Peril & Policy Details & Reinsurance Contract & Risk Reduction \\                                                     
  \midrule    
\endhead

\midrule
\multicolumn{5}{r}{{Continued on next page}} \\
\midrule
\endfoot
\bottomrule
\insertTableNotes         
\endlastfoot     

\begin{tabular}[c]{@{}l@{}}Florida,\\ Florida\\ Hurricane\\ Catastrophe\\ Fund\tnote{1},\\ 1993\end{tabular}                             & Hurricane                                                                                                                                      & \begin{tabular}[c]{@{}l@{}}Property insurance policies must\\ include coverage for hurricane.\\ Policyholders are eligible for\\ premium discounts for installing\\ certain wind resistant features\\ on their homes.\\ Hurricane policies may include\\ a deductible.\end{tabular}                                                                                                                 & \begin{tabular}[c]{@{}l@{}}Insurers are obliged to cede\\ residential property's hurricane\\ risks to the fund. Insurers select\\ a coverage percentage of 45\%,\\ 75\%, or 90\%. A participating\\ insurer’s premium, retention, and\\ coverage limit are based on its\\ total insured values by ZIP code.\end{tabular}                                                                             & \begin{tabular}[c]{@{}l@{}}Retrocession.\\ ILS.\\ +Post-event bonds.\end{tabular}                                                                                                                                                                                                                                                                                                    \\ \hline
\begin{tabular}[c]{@{}l@{}}France,\\ Caisse Centrale\\ de Réassurance\tnote{2},\\ 1946 (natural\\ risks cover\\ from 1982).\end{tabular} & \begin{tabular}[c]{@{}l@{}}Natural events\\ that the\\ government\\ declares\\ disasters,\\ except storms,\\ hail, snow,\\ frost.\end{tabular} & \begin{tabular}[c]{@{}l@{}}Natural disaster coverage is\\ compulsory in all property\\ insurance policies.\\ Rates are set as percentages of\\ the premium of the basic\\ insurance policy.\end{tabular}                                                                                                                                                                                            & \begin{tabular}[c]{@{}l@{}}The Company offers unlimited\\ cover for specific classes of\\ business in the French market.\\ Insurers decide whether to cede\\ risks to CCR or not.\end{tabular}                                                                                                                                                                                                       & \begin{tabular}[c]{@{}l@{}}Guarantee of the French\\ State.\\ Risk Prevention Plans.\end{tabular}                                                                                                                                                                                                                                                                                    \\ \hline
\begin{tabular}[c]{@{}l@{}}Japan,\\ Japanese\\ Earthquake\\ Reinsurance\\ Co.\tnote{3},\\ 1966\end{tabular}                              & \begin{tabular}[c]{@{}l@{}}Earthquakes.\\ Volcanic\\ eruption.\end{tabular}                                                                    & \begin{tabular}[c]{@{}l@{}}Earthquake insurance are\\ compulsorily written with fire\\ policies on residential dwellings\\ and/or personal properties. The\\ amount insured is 30-50\% of\\ the amount provided by the fire\\ policy, but is limited to a fixed\\ maximum. The premium is risk\\ based and set by law. Discounts\\ for earthquake-resistant\\ buildings are available.\end{tabular} & \begin{tabular}[c]{@{}l@{}}JER compensates 100\% of the\\ claim that the insurer paid to the\\ policyholder. Reserves are made\\ up of policyholders' premiums,\\ capital by the government, \\ investment profits from these\\ accumulated liability reserves.\\ If an event occurs, each of JER,\\ non-life insurers and the\\ government pays a claim\\ according to each liability.\end{tabular} & \begin{tabular}[c]{@{}l@{}}JER and non-life insurers pay\\ claims up to 87.1 billion yen\\ per earthquake. The non-life\\ insurers, the government and\\ JER share equally claims for\\ the portion exceeding 87.1\\ up to 153.7 billion yen. The\\ government pays a majority\\ of claims for the portion\\ exceeding 153.7 billion yen.\\ JER also buys retrocession.\end{tabular} \\ \hline
\begin{tabular}[c]{@{}l@{}}UK\\ Flood Re\tnote{4}\\ 2016\end{tabular}                                                                    & Flood                                                                                                                                          & \begin{tabular}[c]{@{}l@{}}Insures decide whether to offer\\ a policy or not and set the price.\end{tabular}                                                                                                                                                                                                                                                                                        & \begin{tabular}[c]{@{}l@{}}Flood Re compensates 100\% of\\ the claim that the insurer paid to\\ the policyholder. Any insurer\\ that offers home insurance in the\\ UK must pay a levy to Flood Re.\\ Insurers can choose to pass the\\ flood risk to Flood Re for a fixed\\ price.\end{tabular}                                                                                                     & \begin{tabular}[c]{@{}l@{}}Government's commitment\\ to risk mitigation.\\ Flood Re buys its own\\ reinsurance programme every\\ three years to cover losses of\\ up to £2.2bn per annum.\end{tabular}                                                                                                                                                                               \\ \hline

\end{longtable}
\end{ThreePartTable}

\end{landscape}

\end{document}